\documentclass[conference]{IEEEtran}
\RequirePackage[rightcaption]{sidecap}
\usepackage{cite}
\usepackage{amsmath,amssymb,amsfonts}
\usepackage{algorithmic}
\usepackage{graphicx}
\usepackage{subcaption}
\usepackage{fancybox}
\usepackage{url}
\usepackage{textcomp}
\usepackage[switch]{lineno}
\usepackage{xcolor}
\usepackage[colorlinks=true, allcolors=blue]{hyperref}
\def\BibTeX{{\rm B\kern-.05em{\sc i\kern-.025em b}\kern-.08em
    T\kern-.1667em\lower.7ex\hbox{E}\kern-.125emX}}
\usepackage{datetime}
\yyyymmdddate

\usepackage{multirow}

\def\frameworktwo{\texttt{RETINA}}
\def\expandtwo{{\bf Ret}weeter {\bf I}dentifier {\bf N}etwork with Exogenous {\bf A}ttention}
\newcommand{\framework}{\frameworktwo}
\newcommand{\frameworkexpand}{\expandtwo}

\begin{document}

\title{Hate is the New Infodemic: A Topic-aware Modeling of Hate Speech Diffusion on Twitter}

\author{
\IEEEauthorblockN{ Sarah Masud\textsuperscript{$\dagger$}, Subhabrata Dutta\textsuperscript{$\star$}, Sakshi Makkar\textsuperscript{$\dagger$}, Chhavi Jain\textsuperscript{$\dagger$}, Vikram Goyal\textsuperscript{$\dagger$}, \\Amitava Das\textsuperscript{$\ddagger$}, Tanmoy Chakraborty\textsuperscript{$\dagger$}}
\IEEEauthorblockA{%\textit{dept. name of organization (of Aff.)} \\
\textit{\textsuperscript{$\star$}Jadavpur University, Kolkata, India; \textsuperscript{$\dagger$}IIIT-Delhi, New Delhi, India; \textsuperscript{$\ddagger$}Wipro AI, Bangalore, India}\\
subha0009@gmail.com; \{sarahm, sakshi18013, chhavi19117, vikram, tanmoy\}@iiitd.ac.in; amitava.das2@wipro.com}}
\maketitle

\begin{abstract}
Online hate speech, particularly over microblogging platforms like Twitter, has emerged as arguably the most severe issue of the past decade. Several countries have reported a steep rise in hate crimes infuriated by malicious hate campaigns. While the detection of hate speech is one of the emerging research areas, the generation and spread of topic-dependent hate in the information network remain under-explored. In this work, we focus on exploring user behavior, which triggers the genesis of hate speech on Twitter and how it diffuses via retweets. We crawl a large-scale dataset of tweets, retweets, user activity history, and follower networks, comprising over $161$ million tweets from more than $41$ million unique users. We also collect over $600k$ contemporary news articles published online. We characterize different signals of information that govern these dynamics. Our analyses differentiate the diffusion dynamics in the presence of hate from usual information diffusion. This motivates us to formulate the modeling problem in a {\em topic-aware setting} with real-world knowledge. For predicting the initiation of hate speech for any given hashtag, we propose multiple feature-rich models, with the best performing one achieving a macro F1 score of $0.65$. Meanwhile, to predict the retweet dynamics on Twitter, we propose \framework, a novel neural architecture that incorporates exogenous influence using scaled dot-product attention. \framework\ achieves a macro F1-score of $0.85$, outperforming multiple state-of-the-art models. Our analysis reveals the superlative power of \framework\ to predict the retweet dynamics of hateful content compared to the existing diffusion models. 
\end{abstract}

\section{Introduction}
\label{sec:intro}
For the past half-a-decade, in synergy with the socio-political and cultural rupture worldwide, online hate speech has manifested as one of the most challenging issues of this century transcending beyond the cyberspace. Many hate crimes against minority and backward communities have been directly linked with hateful campaigns circulated over Facebook, Twitter, Gab, and many other online platforms \cite{united2018report, muller2019fanning}. Online social media has provided an unforeseen speed of information spread, aided by the fact that the power of content generation is handed to every user of these platforms. Extremists have exploited this phenomenon to disseminate hate campaigns to a degree where manual monitoring is too costly, if not impossible.

Thankfully, the research community has been observing a spike of works related to online hate speech, with a vast majority of them focusing on the problem of automatic detection of hate from online text \cite{fortuna2018survey}. However, as Ross et al. \cite{DBLP:journals/corr/RossRCCKW17} pointed it out, even manual identification of hate speech comes with ambiguity due to the differences in the definition
of hate. Also, an important signal of hate speech is the presence of specific words/phrases, which vary significantly across topics/domains. Tracking such a diverse socio-linguistic phenomenon in real-time is impossible for automated, large-scale platforms.  

 An alternative approach can be to track potential groups of users who have a history of spreading hate. As Matthew et al. \cite{mathew2019spread} suggested, such users are often a very small fraction of the total users but generate a sizeable portion of the content. Moreover, the severity of hate speech lies in the degree of its spread, and an early prediction of the diffusion dynamics may help combat online hate speech to a new extent altogether. However, a tiny fraction of the existing literature seeks to explore the problem quantitatively. Matthew et al. \cite{mathew2019spread} put up an insightful foundation for this problem by analyzing the dynamics of hate diffusion in Gab\footnote{\protect\url{https://gab.com/}}. However, they do not tackle the problem of modeling the diffusion and restrict themselves to identifying different characteristics of hate speech in Gab.

{\bf Hate speech on Twitter:} Twitter, as one of the largest micro-blogging platforms with a worldwide user base, has a long history of accommodating hate speech, cyberbullying, and toxic behavior. Recently, it has come hard at such contents multiple times\footnote{\protect\url{https://blog.twitter.com/en_us/topics/company/2019/hatefulconductupdate.html}}, and a certain fraction of hateful tweets are often removed upon identification. However, a large majority of such tweets still circumvent Twitter's filtering. In this work, we choose to focus on the dynamics of hate speech on Twitter mainly due to two reasons: (i) the wide-spread usage of Twitter compared to other platforms provides scope to grasp the hate diffusion dynamics in a more realistic manifestation, and (ii) understanding how hate speech emerges and spreads even in the presence of some top-down checking measures, compared to unmoderated platforms like Gab. 
% \footnote{\protect\url{https://www.bbc.com/news/technology-42376546}}

\begin{figure}[!t]
    \centering
    \includegraphics[width=\columnwidth]{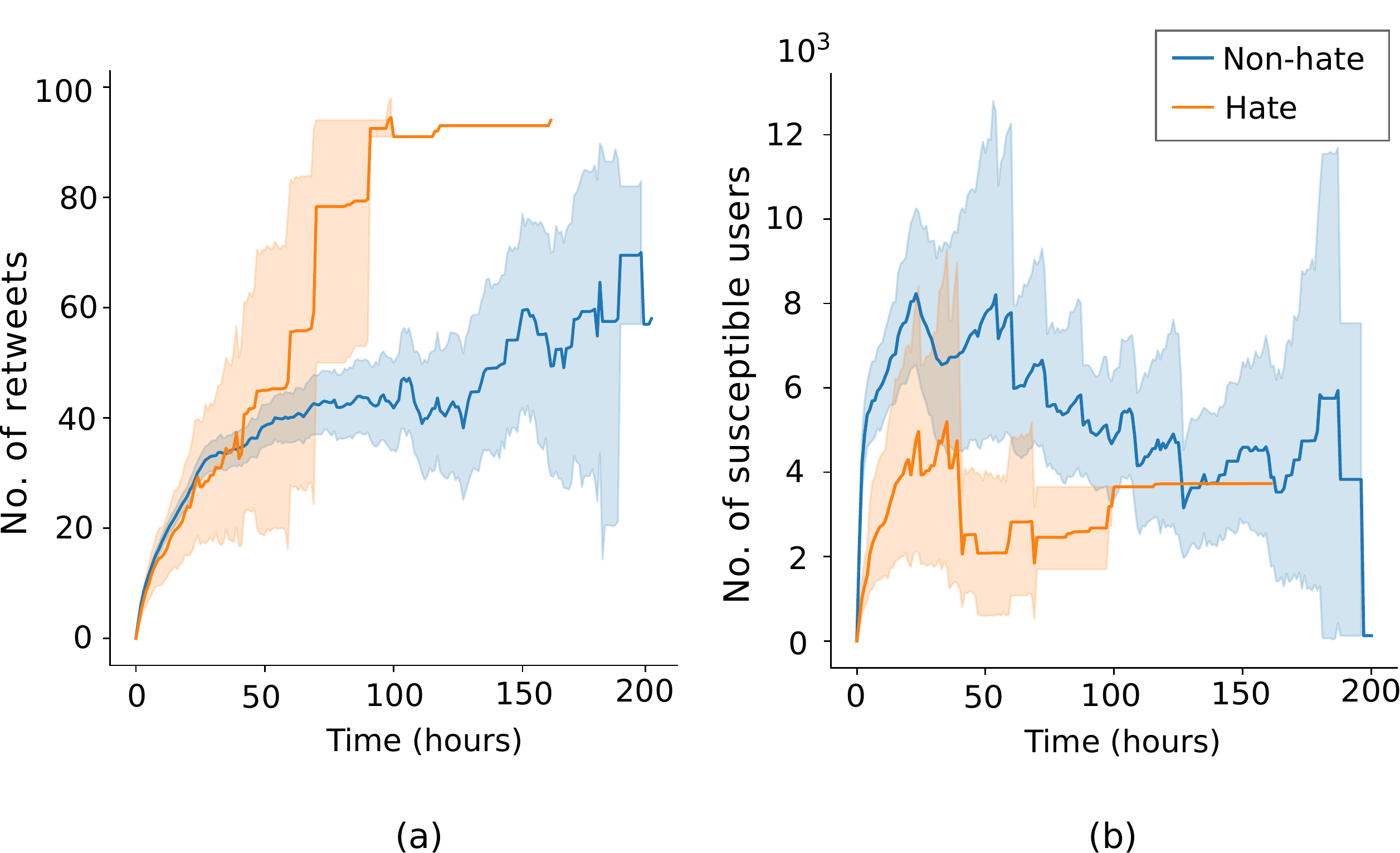}
    
   \caption{Plot (a) shows the growth of retweet cascades for hateful and non-hate tweets (solid lines and shaded regions signify the average over the dataset and confidence of count, respectively). Analogously, plot (b) depicts the temporal change of susceptible users over time.}
    \label{fig:cascade_susceptible}
    \vspace{-5mm}
\end{figure}

{\bf Diffusion patterns of hate vs. non-hate on Twitter:} Hate speech is often characterized by the formation of {\em echo-chambers}, i.e., only a small group of people engaging with such contents repeatedly. In Figure~\ref{fig:cascade_susceptible}, we compare the temporal diffusion dynamics of hateful vs. non-hate tweets (see Sections \ref{subsec:data} and \ref{subsec:hate_detect} for the details of our dataset and hate detection methods, respectively). Following the standard information diffusion terminology, the set of {\em susceptible} nodes at any time instance of the spread is defined by all such nodes which have been exposed to the information (followers of those who have posted/retweeted the tweet) up to that instant but did not participate in spreading (did not retweet/like/comment). While hateful tweets are retweeted in a significantly higher magnitude compared to non-hateful ones (see Figure~\ref{fig:cascade_susceptible}(a)), they tend to create lesser number of susceptible users over time (see Figure~\ref{fig:cascade_susceptible}(b)). This is directly linked to two major phenomena: primarily, one can relate this to the formation of hate echo-chambers -- hateful contents are distributed among a well-connected set of users. Secondarily, as we define susceptibility in terms of follower relations,  hateful contents, therefore, might have been diffusing among connections beyond the follow network -- through paid promotion, etc. Also one can observe the differences in early growth for the two types of information; while hateful tweets acquire most of their retweets and susceptible nodes in a very short time and stall, later on, non-hateful ones tend to maintain the spread, though at a lower rate, for a longer time. This characteristic can again be linked to organized spreaders of hate who tend to disseminate hate as early as possible.

\begin{figure}[!t]
    \centering
    \includegraphics[width=0.8\columnwidth]{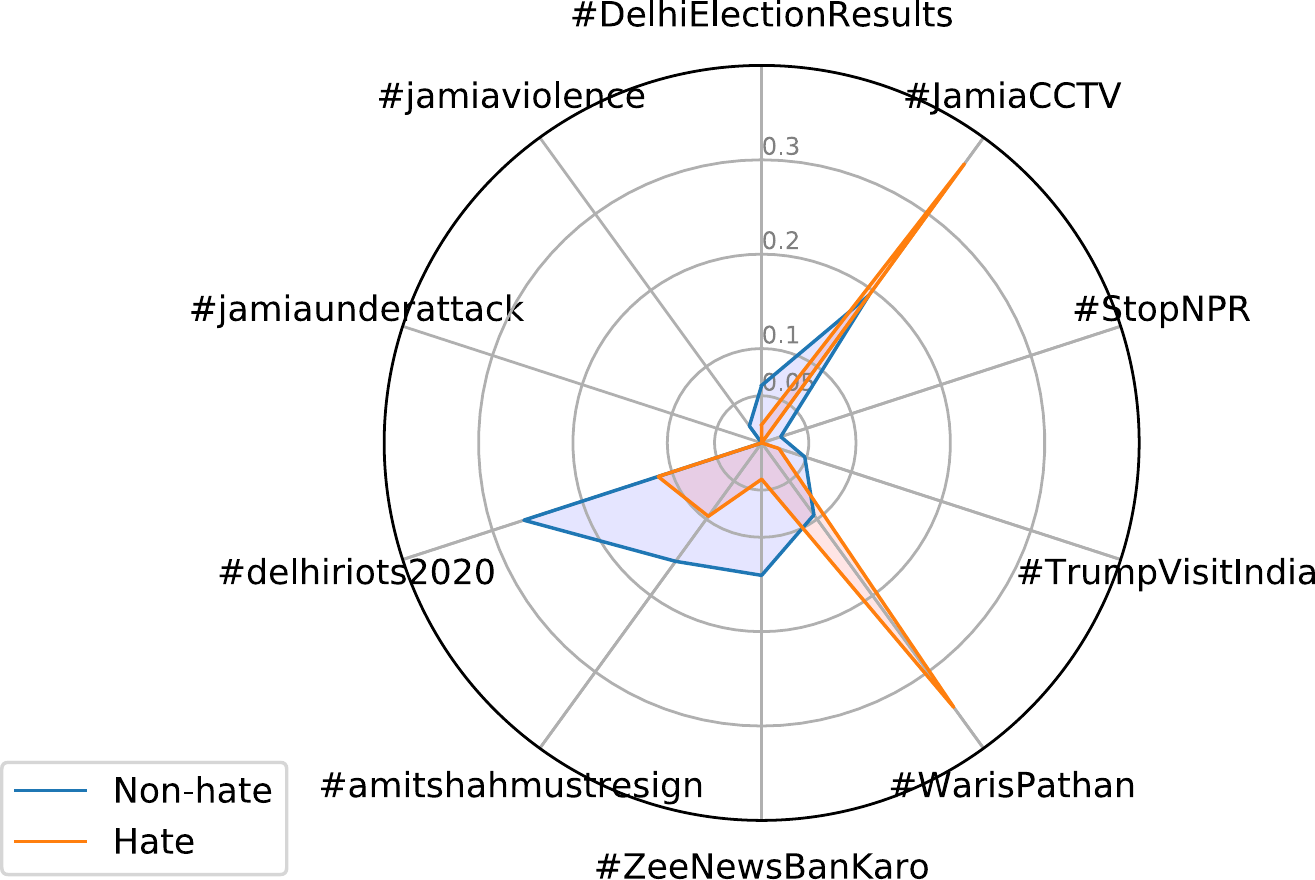}
    \caption{Distribution of hateful vs non-hate tweets (on a scale $0$ to $1$) for a selected number of hashtags.}
    \label{fig:topic_hate_dist}
\end{figure}

\begin{figure}
    \centering
    \includegraphics[width=\columnwidth]{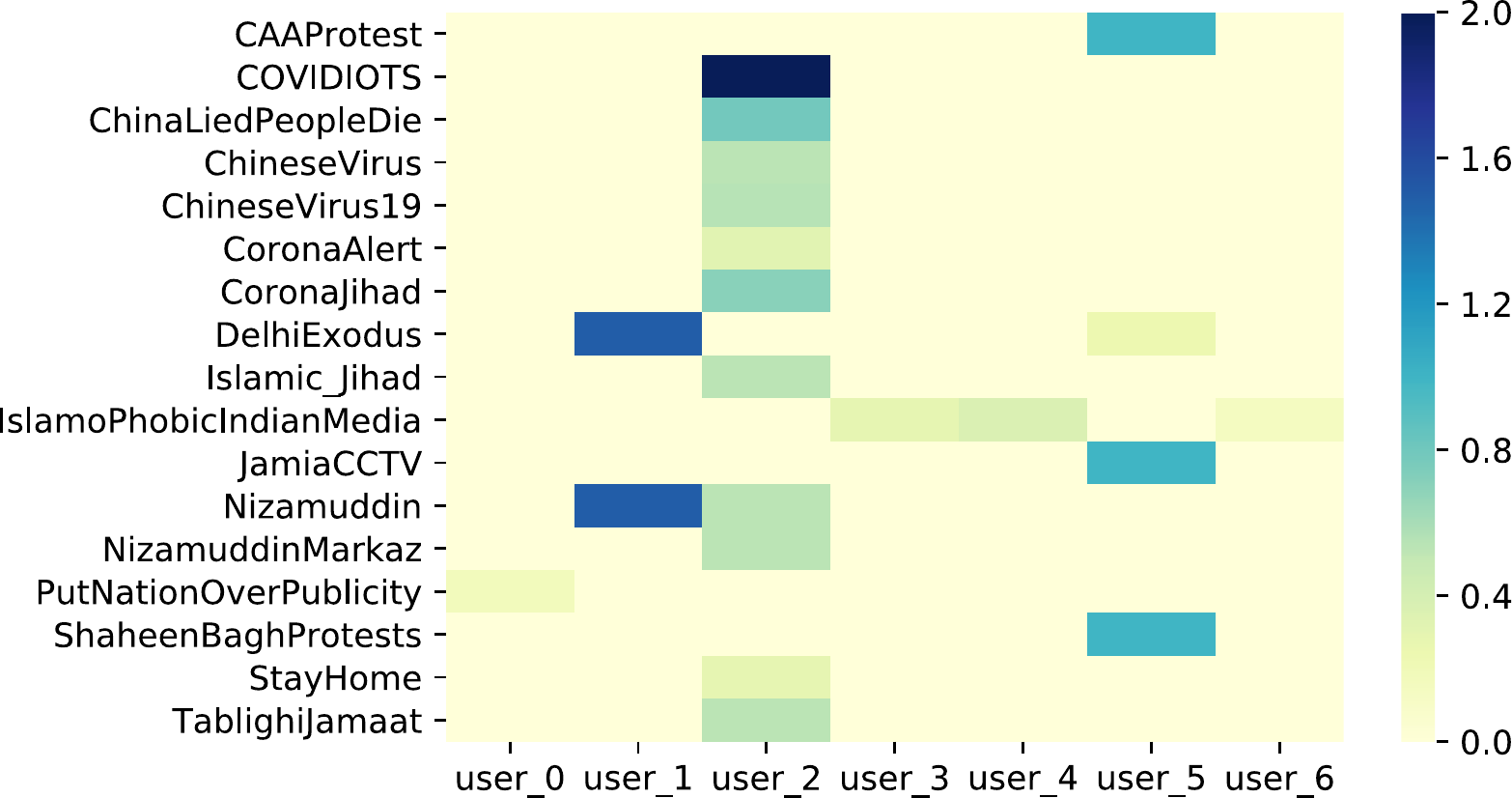}
    \caption{Distribution of hatefulness expressed by a selected set of users for different hashtags. The color of a cell corresponds to a user, and a hashtag signifies the ratio of hateful to non-hate tweets posted by that user using that specific hashtag. }
    \label{fig:user-topic-hate}
    \vspace{-5mm}
\end{figure}

{\bf Topic-dependence of Twitter hate:} Hateful contents show strong topic-affinity:  topics related to politics and social issues, for example, incur much more hateful content compared to sports or science. Hashtags in Twitter provide an overall mapping for tweets to topics of discussion. As shown in Figure~\ref{fig:topic_hate_dist}, the degree of hateful content varies significantly for different hashtags. Even when different hashtags share a common theme (such as {\em topic of discussion \#jamiaunderattack, \#jamiaviolence} and {\em \#jamiaCCTV}), they may still incur a different degree of hate. Previous studies \cite{mathew2019spread} tend to denote users as hate-preachers irrespective of the topic of discussion. However, as evident in Figure~\ref{fig:user-topic-hate}, the degree of hatefulness expressed by a user is dependent on the topic as well. For example, while some users resort to hate speech concerning COVID-19 and China, others focus on topics around the protests against the Citizenship Amendment Act in India.

{\bf Exogenous driving forces:} With the increasing entanglement of virtual and real social processes, it is only natural that events happening outside the social media platforms tend to shape the platform's discourse. Though a small number of existing studies attempt to inquire into such inter-dependencies \cite{Dutta2020DeepEA, Exo_MIT}, the findings are substantially motivating in problems related to modeling information diffusion and user engagement in Twitter and other platforms. In the case of hate speech, exogenous signals offer even more crucial attributes to look into, which is {\em global context}. For both detecting and predicting the spread of hate speech over short tweets, the knowledge of context is likely to play a decisive role 
% SHORT URL NOT WORKING: (e.g., there has been $\sim 9$-times rise in hate tweets aimed at Chinese people after the COVID-19 broke out\footnote{\protect\url{https://shorturl.at/dJPS0}}).

{\bf Present work:} Based on the findings of the existing literature and the analysis we presented above, here we attempt to model the dynamics of hate speech spread on Twitter. We separate the process of spread as the {\em hate generation} (asking for who will start a hate campaign) and {\em retweet diffusion of hate} (who will spread an already started hate campaign via retweeting). To the best of our knowledge, this is the {\bf very first attempt to delve into the predictive modeling of online hate speech}. Our contributions can be summarized as follows:
\begin{enumerate}
    \item We formalize the dynamics of hate generation and retweet spread on Twitter subsuming, the activity history of each user and signals propagated by the localized structural properties of the information network of Twitter induced by follower connections as well as global endogenous and exogenous signals (events happening inside and outside of Twitter) (See Section~\ref{sec:abstract_model}).
    
    \item We present a large dataset of tweets, retweets, user activity history, and the information network of Twitter covering versatile hashtags, which made to trend very recently. We manually annotate a significant subset of the data for hate speech. We also provide a corpus of contemporary news articles published online (see Section~\ref{subsec:data} for more details).
    
    \item We unsheathe rich set of features manifesting the signals mentioned above to design multiple prediction frameworks which forecast, given a user and a contemporary hashtag, whether the user will write a hateful post or not (Section~\ref{sec:hate_generate}). We provide an in-depth feature ablation and ensemble methods to analyze our proposed models' predictive capability, with the best performing one resulting in a macro F1-score of $\mathbf{0.65}$.
    
    \item We propose \framework\ (\frameworkexpand), a neural architecture to predict potential retweeters given a tweet (Section~\ref{subsec:framework}). \framework\ encompasses an attention mechanism which dictates the prediction of retweeters based on a stream of contemporary news articles published online. Features representing hateful behavior encoded within the given tweet as well as the activity history of the users further help \framework\ to achieve a macro F1-score of $\mathbf{0.85}$, significantly outperforming several state-of-the-art retweet prediction models. 
\vspace{4mm}
\end{enumerate}

\noindent\shadowbox{\begin{minipage}[t]{.95\columnwidth}
{\bf Reproducibility:} We have made public our datasets and code along with the necessary instructions and parameters, available at \href{https://github.com/LCS2-IIITD/RETINA}{https://github.com/LCS2-IIITD/RETINA}.
\end{minipage}}

\section{Related Work}
{\bf Hate speech detection.} In recent years, the research community has been keenly interested in better understanding, detection, and combating hate speech on online media. Starting with the basic feature-engineered Logistic Regression models\cite{waseem-hovy-2016-hateful, Davidson2017AutomatedHS}  to the latest ones employing neural architectures \cite{PinkeshHate}, a variety of automatic online hate speech detection models have been proposed across languages \cite{Hate_Fortuna_Port}. To determine the hateful text, most of these models utilize a static-lexicon based approach and consider each post/comment in isolation. With lack of context (both in the form of individual's prior indulgence in the offense and the current world view), the models trained on previous trends perform poorly on new datasets. While linguistic and contextual features are essential factors of a hateful message, the destructive power of hate speech lies in its ability to spread across the network. However, only recently have researchers started using network-level information for hate speech detection \cite{ghosh-chowdhury-etal-2019-arhnet}, \cite{fehn-unsvag-gamback-2018-effects}.  Rathpise and Adji \cite{rathpisey2019handling} proposed methods to handle class imbalance in hate speech classification. A recent work showed how the anti-social behavior on social media during COVID-19 led to the spread of hate speech. Awal et al. \cite{awal2020analyzing} coined the term,
`disability hate speech' and showed its social, cultural and political contexts. Ziems et al.  \cite{ziems2020racism} explained how COVID-19 tweets increased racism, hate, and xenophobia in social media.

While our work does not involve building a new hate speech detection model, yet hate detection underpins any work on hate diffusion in the first place. Inspired by existing research, we also incorporate hate lexicons as a feature for the diffusion model. The lexicon is curated from multiple sources and manually pruned to suit the Indian context \cite{kapoor2018mind}. Meanwhile, to overcome the problem of context, we utilize the timeline of a user to determine her propensity towards hate speech.

{\bf Information diffusion and microscopic prediction.}
Predicting the spread of information on online platforms is crucial in understanding the network dynamics with applications in marketing campaigns, rumor spreading/stalling, route optimization, etc. The latest in the family of diffusion being the CHASSIS \cite{CHASSIS} model. On the other end of the spectrum, the SIR model \cite{SIR_Original} effectively captures the presence of R (Recovered) nodes in the system, which are no longer active due to information fatigue\footnote{\protect{\url{http://paginaspersonales.deusto.es/abaitua/konzeptu/fatiga.htm}}}. Even though limited in scope, the SIR model serves as an essential baseline for all diffusion models. 

Among other techniques, a host of studies employ social media data for both macroscopic (size and popularity) and microscopic (next user(s) in the information cascade) prediction. While highly popular, both DeepCas \cite{DeepCas} and DeepHawkes \cite{DeepHawkes} focus only on the size of the overall cascade. Similarly, Khosla et al. \cite{AdityaPopImage} utilized social cues to determine the popularity of an image on Flickr. While Independent Cascade (IC) based embedding models \cite{NodeIC_Trans,Gao2017ANE} led the initial work in supervised learning based microscopic cascade prediction; they failed to capture the cascade's temporal history (either directly or indirectly). Meanwhile,  Yang et al. \cite{Yang2018NeuralDM} presented a neural diffusion model for microscopic prediction, which employs recurrent neural architecture to capture the history of the cascade. These models focus on predicting the next user in the cascade from a host of potential candidates. In this regard, TopoLSTM \cite{TopoLSTM} considers only the previously seen nodes in any cascade as the next candidate without using timestamps as a feature. This approximation works well under limited availability of network information and the absence of cascade metadata. Meanwhile, FOREST \cite{Yang2019MultiscaleID} considers all the users in the global graph (irrespective of one-hop) as potential users, employing a time-window based approach. Work by Wang et al. \cite{Wang2019HierarchicalDA} lies midway of TopoLSTM and FOREST, in that it does not consider any external global graph as input, but employs a temporal, two-level attention mechanism to predict the next node in the cascade. Zhou et al. \cite{Zhou2020ASO}  compiled a detailed outline of recent advances in cascade prediction.

Compared to the models discussed above for microscopic cascade prediction, which aim to answer who will be the next participant in the cascade, our work aims to determine whether a follower of a user will retweet (participate in the cascade) or not.  This converts our use case into a binary classification problem, and adds negative sampling (in the form on inactive nodes), taking the proposed model closer to real-world scenario consisting of active and passive social media users.

The spread of hate and exploratory analysis by Mathew et al. \cite{mathew2019spread} revealed exciting characteristics of the breadth and depth of hate vs. non-hate diffusion. However, their methodology separates the non-haters from haters and studies the diffusion of two cascades independently. Real-world interactions are more convoluted with the same communication thread containing hateful, counter-hateful, and non-hateful comments. Thus, independent diffusion studies, while adequate at the exploratory analysis of hate, cannot be directly extrapolated for predictive analysis of hate diffusion. The need is a model that captures the hate signals at the user and/or group level. By taking into account the user's timeline and his/her network traits, we aim to capture more holistic hate markers.

{\bf Exogenous influence.}
As early as 2012, Myers et al. \cite{Exo_MIT} exposed that external stimuli drive one-third of the information diffusion on Twitter. Later, Hu et al. \cite{Hu2015PredictingUE} proposed a model for predicting user engagement on Twitter that is factored by user engagement in 600 real-world events. From employing world news data for enhancing language models \cite{Wenzek2020CCNetEH} to boosting the impact of online advertisement campaigns \cite{Event_Triggered}, exogenous influence has been successfully applied in a wide variety of tasks. Concerning social media discourse, both De et al. \cite{AbirIITK} in opinion mining and Dutta et al. \cite{Dutta2020DeepEA} in chatter prediction corroborated the superiority of models that consider exogenous signals.

Since our data on Twitter was collected based on trending Indian hashtags, it becomes crucial to model exogenous signals, some of which may have triggered a trend in the first place. While a one-to-one mapping of news keywords to trending keywords is challenging to obtain, we collate the most recent (time-window) news w.r.t to a source tweet as our ground-truth. {\em To our knowledge, this is the first retweet prediction model to consider external influence. }

\section{Topic-aware Generation and Diffusion of Content over Information Networks}
\label{sec:abstract_model}

\begin{table}
\caption{Important notations and denotations.}
    \centering
    \small
    \begin{tabular}{|c|c|}
    \hline
        Notation & Denotation \\
    \hline
        $\mathcal{G}$ & The information network\\
        $\mathcal{H}_{i,t}$ & Activity history of user $u_i$ up to time $t$\\
        $\mathcal{S}^\text{ex}$ & Exogenous influence\\
        $\mathcal{S}^\text{en}$ & Endogenous influence\\
        $\mathcal{S}^P_i$ & Peer influence on $u_i$\\
        $\mathcal{T}$ & Topic (hashtag)\\
        $P^{u_i}$, $P^{u_i}_j$ & Probability of $u_i$ retweeting (static vs. $j^\text{th}$ interval) \\
        $\mathbf{X}^T$, $\mathbf{X}^N$ & Feature tensors for tweet and news\\
        $\mathbf{X}^{T,N}$ & Output from exogenous attention\\ 
    \hline
    \end{tabular}
    \label{tab:not_denot}
    \vspace{-5mm}
\end{table}

An information network of Twitter can be defined as a directed graph $\mathcal{G} = \{\mathcal{U}, \mathcal{E}\}$, where every user corresponds to a unique node $u_i\in \mathcal{U}$, and there exists an ordered pair $(u_i, u_j)\in \mathcal{E}$ if and only if the user corresponding to $u_j$ follows user $u_i$. (Table \ref{tab:not_denot} summarizes important notations and denotations.) Typically, the visible information network of Twitter does not associate the follow relation with any further attributes, therefore any two edges in $\mathcal{E}$ are indistinguishable from each other. We associate unit weight to every $e\in \mathcal{E}$.

Every user in the network acts as an agent of content generation (tweeting) and diffusion (retweeting). For every user $u_i$ at time $t_0$, we associate an activity history $\mathcal{H}_{i, t_0} = \{\tau(t)|t\leq t_0\}$, where $\tau(t)$ signifies a tweet posted (or retweeted) by $u_i$ at time $t$. 

The information received by user $u_i$ has three different sources: (a) {\em Peer signals} ($\mathcal{S}^P_i$): The information network $\mathcal{G}$ governs the flow of information from node to node such that any tweet posted by $u_i$ is visible to every user $u_j$ if $(u_i, u_j)\in \mathcal{E}$; (b) {\em Non-peer endogenous signals} ($\mathcal{S}^{\text{en}}$): Trending hashtags, promoted contents, etc. that show up on the user's feed even in the absence of peer connection; (c) {\em Exogenous signals} ($\mathcal{S}^{\text{ex}}$): Apart from the Twitter feed, every user interacts with the external world-events directly (as a participant) or indirectly (via news, blogs, etc.). 

{\bf Hate generation}. The problem of modeling hate generation can be formulated as assigning a probability with each user that signifies their likelihood to post a hateful tweet. With our hypothesis of hateful behavior being a topic-dependent phenomenon, we formalize the modeling problem as learning the parametric function, $f_1:{\rm I\!R}^{d}\rightarrow (0, 1)$ such that,
\begin{equation}
\label{Eq:tweet}
    P(u_i|\mathcal{T}) = f_1(\mathcal{S}^{\text{en}}, \mathcal{S}^{\text{ex}}, \mathcal{H}_{i, t}, \mathcal{T}|\theta_1)
\end{equation}
where $\mathcal{T}$ is a given topic, $t$ is the instance up to which we obtain the observable history of $u_i$, $d$ is the dimensionality of the input feature space, and $\theta_1$ is the set of learnable parameters. Though ideally $P(u_i|\mathcal{T})$ should be dependent on $\mathcal{S}^P_i$ as well, the complete follower network for Twitter remains mostly unavailable due to account settings, privacy constraints, inefficient crawling, etc.

{\bf Hate diffusion}. As already stated, we characterize diffusion as the dynamic process of retweeting in our context. Given a tweet $\tau(t_0)$ posted by some user $u_i$, we formulate the problem as predicting the potential retweeters within the interval $[t_0, t_0+\Delta t]$. Assuming the probability density of a user $u_j$ retweeting $\tau$ at time $t$ to be $p(t)$, then retweet prediction problem translates to learning the parametric function $f_2:{\rm I\!R}^{d}\rightarrow (0, 1)$ such that,
\begin{equation}
\label{Eq:retweet}
    \int^{t_0+\Delta t}_{t_0}p(t)dt = f_2(\mathcal{S}^P_j, \mathcal{S}^{\text{en}}_j, \mathcal{S}^{\text{ex}}_j, \mathcal{H}_{j, t}, \tau|\theta_2)
\end{equation}

Eq.~\ref{Eq:retweet} is the general form of a parametric equation describing retweet prediction. In our setting, the signal components $\mathcal{S}^P_j$, $\mathcal{H}_{j, t}$, and the features representing the tweet $\tau$ incorporates the knowledge of hatefulness. Henceforth, we call $\tau$ the {\em root tweet} and $u_i$ the {\em root user}. It is to be noted that, the features representing the peer, non-peer endogenous, and exogenous signals in Eq.~\ref{Eq:tweet} and \ref{Eq:retweet} may differ due to the difference in problem setting. 

{\bf Beyond organic diffusion.} The task of identifying potential retweeters of a post on Twitter is not straightforward. In retrospect, the event of a user retweeting a tweet implies that the user must have been an audience of the tweet at some point of time (similar to `susceptible' nodes of contagion spread in the SIR/SIS models \cite{SIR_Original},\cite{SIS_Original}). For any user, if at least one of his/her followees engages with the retweet cascade, then the subject user becomes susceptible. That is, in an {\em organic} diffusion, between any two users $u_i, u_j$ there exists a finite path $\langle u_i, u_{i+1} \dots, u_{j} \rangle$ in $\mathcal{G}$ such that each user (except $u_i$) in this path is a retweeter of the tweet by $u_i$. However, due to account privacy etc., one or more nodes within this path may not be visible. Moreover, contents promoted by Twitter, trending topics, content searched by users independently may diffuse alongside their organic diffusion path. Searching for such retweeters is impossible without explicit knowledge of these phenomena. Hence, we primarily restrict our retweet prediction to the organic diffusion, though we experiment with retweeters not in the visibly organic diffusion cascade to see how our models handle such cases.

\begin{figure*}[!t]
    \centering
    \includegraphics[width=\textwidth]{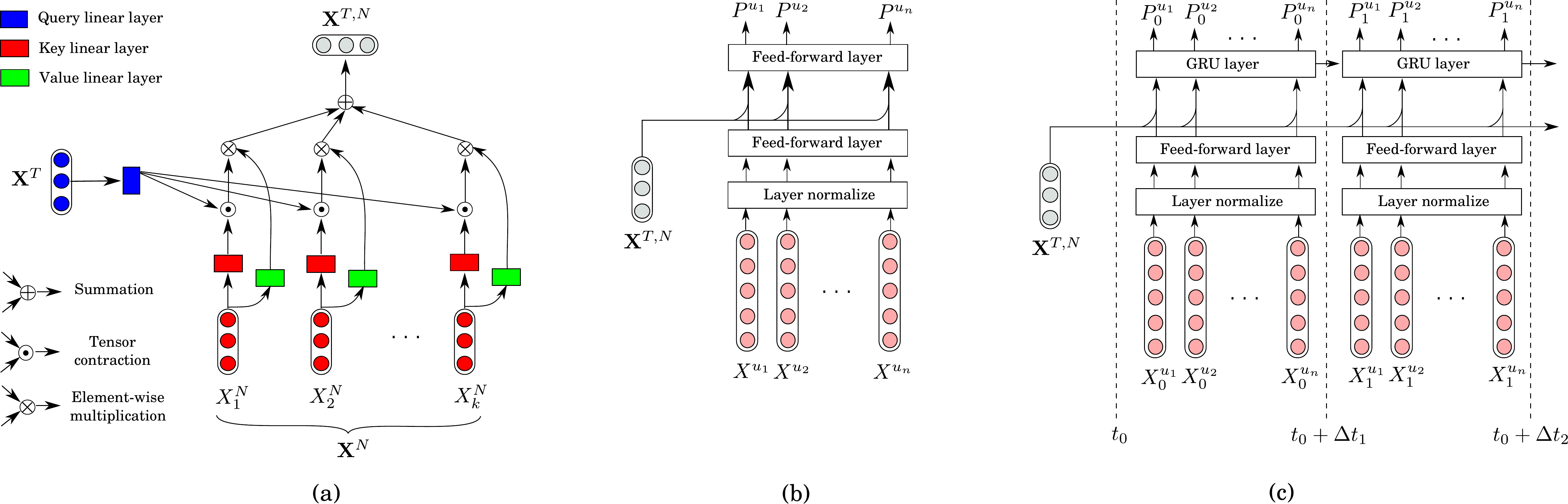}
    \caption{{\bf Design of different components of \framework} -- (a) {\em Exogenous attention}: Key and Value linear layers (blue) are applied on each element of the news feature sequence $\mathbf{X}^N$, while the Query linear layer (red) is applied on the tweet feature $\mathbf{X}^T$. The attention weights computed for each news feature vector by contracting the query and key tensors along feature axis (dot product) are then applied to the value tensors and summed over the sequence axis to produce the `attended' output, $\mathbf{X}^{T,N}$. (b) {\em Static prediction of retweeters}: To predict whether $u_j$ will retweet, the input feature $X^{u_j}$ is normalized and passed through a feed-forward layer, concatenated with $\mathbf{X}^{T,N}$, and another feed-forward layer is applied to predict the retweeting probability $P^{u_j}$. (c) {\em Dynamic retweet prediction}: In this case, \framework\ predicts the user retweet probability for consecutive time intervals, and instead of the last feed-forward layer used in the static prediction, we use a GRU layer.}
    \label{fig:framework}
    \vspace{-5mm}
\end{figure*}

\section{Modeling Hate Generation}
\label{sec:hate_generate}
To realize Eq.~\ref{Eq:tweet}, we signify topics as individual hashtags. We rely purely on manually engineered features for this task so that rigorous ablation study and analysis produce explainable knowledge regarding this novel problem. The extracted features instantiate different input components of $f_1$ in Eq.~\ref{Eq:tweet}. We formulate this task in a static manner, i.e., assuming that we are predicting at an instance $t_0$, we want to predict the probability of the user posting a hateful tweet within $[t_0, \infty]$. While training and evaluating, we set $t_0$ to be right before the actual tweeting time of the user.

\subsection{User history-based features} 
\label{subsec:user-history}
The activity history of user $u_i$, signified by $\mathcal{H}_{i,t}$ is substantiated by the following features: 

$\bullet$ We use unigram and bigram features weighted by tf-idf values from $30$ most recent tweets posted by $u_i$ to capture its recent topical interest. To reduce the dimensionality of the feature space, we keep the top $300$ features sorted by their idf values.

$\bullet$ To capture the history of hate generation by $u_i$, we compute two different features her most recent $30$ tweets: (i) ratio of hateful vs. non-hate tweets and (ii) a hate lexicon vector $HL=\{h_i|h_i\in{\rm I\!I}^{+}\text{ and } i=1,\dots, \lvert \mathbf{H} \rvert\}$, where $\mathbf{H}$ is a dictionary of hate words, and $h_i$ is the frequency of the $i^{\text{th}}$ lexicon from $\mathbf{H}$ among the tweet history.

$\bullet$ Users who receive more attention from fellow users for hate propagation are more likely to generate hate. Therefore, we take the ratio of retweets of previous hateful tweets to non-hateful ones by $u_i$. We also take the ratio of total number of retweets on hateful and non-hateful tweets of $u_i$ .

$\bullet$ Follower count and date of account creation of $u_i$.

$\bullet$ Number of topics (hashtags)  $u_i$ has tweeted on up to $t$.

\subsection{Topic (hashtag)-oriented feature}
We compute Doc2Vec \cite{le2014distributed} representations of the tweets, along with the hashtags present in them as individual tokens. We then compute the average cosine similarity between the user's recent tweets and the word vector representation of the hashtag, this serves as the topical relatedness of the user towards the given hashtag.
\subsection{Non-peer endogenous features}
To incorporate the information of trending topics over Twitter, we supply the model with a binary vector representing the top $50$ trending hashtags for the day the tweet is posted.
\subsection{Exogenous feature}
\label{subsec:exogen_feature}
We compute the average tf-idf vector for the $60$ most recent news headlines from our corpus posted before the  time of the tweet. Again we select the top $300$ features. 

Using the above features, we implement six different classification models(and their variants). Details of the models are provided in Section \ref{subsec:task1_details}.

\section{Retweet Prediction}
\label{sec:retweet_prediction}

While realizing Eq.~\ref{Eq:retweet} for retweeter prediction, we formulate the task in two different settings: the {\em static retweeter prediction} task, where $t_0$ is fixed, and $\Delta t$ is $\infty$ (i.e., all the retweeters irrespective of their retweet time) and the {\em dynamic retweeter prediction} task where we predict on successive time intervals.

For these tasks, we rely on features both  designed manually as well as extracted using unsupervised/self-supervised manner. 

\subsection{Feature selection} 
\label{subsec:feature_for_retweet}
For the task of retweet prediction, we extract features representing the root tweet itself, as well as the signals of Eq. \ref{Eq:retweet} corresponding to each user $u_i$ (for which we predict the possibility of retweeting). Henceforth, we indicate the root user by $u_0$.

Here, we incorporate $\mathcal{S}^P_i$ using two different features: shortest path length from $u_0$ to $u_i$ in $\mathcal{G}$, and number of times $u_i$ has retweeted tweets by $u_0$. All the features representing $\mathcal{H}_{i,t}$ and $\mathcal{S}^{en}$ remain same as described in Section \ref{sec:hate_generate}.

We incorporate two sets of features representing the root tweet $\tau$: the hate lexicon vector similar to Section \ref{subsec:user-history} and top $300$. We varied the size of features from $100$ to $1000$, and the best combination was found to be $300$.

For the retweet prediction task, we incorporate the exogenous signal in two different methods. To implement the attention mechanism of \framework, we use a Doc2Vec representations of the news articles as well as the root tweet. For the rest of the models, we use the same feature set as Section \ref{subsec:exogen_feature}.

% \begin{itemize}
%     \item A unigram and bi-gram vector representing the root tweet (top $300$ features weighted by tf-idf values).
%     \item Unigram and bi-gram feature representing the recent tweeting history of the users (similar to Section~\ref{subsec:user-history}).
%     \item Ratio of hate to non-hate tweets of the user.
%     \item  Ratio of number of retweets on hate tweet to the number of retweets on non-hate tweet.
%     \item Follower count of the user
%     \item Date of account creation of the user.
%     \item Hate lexicon vector similar to Section~\ref{subsec:user-history}.
%     \item Length of the shortest path from the root user to the sample user along $\mathcal{G}$.
%     \item Number of times this user has retweeted the root user.
%     \item Number of topics the user has tweeted.
    
% \end{itemize}

\subsection{Design of \framework}
\label{subsec:framework}
Guided by Eq.~\ref{Eq:retweet}, \framework\ exploits the features described in Section~\ref{subsec:feature_for_retweet} for both static and dynamic prediction of retweeters.

{\bf Exogenous attention.} To incorporate external information as an assisting signal to model diffusion, we use a variation of {\em scaled dot product attention} \cite{vaswani2017attention} in \framework\ (see Figure~\ref{fig:framework}). Given the feature representation of the tweet $\mathbf{X}^T$ and news feature sequence $\mathbf{X}^N=\{X^N_1, X^N_2, \dots, X^N_k\}$, we compute three tensors $\mathbf{Q}^T$, $\mathbf{K}^N$, and $\mathbf{V}^N$, respectively as follows:
\begin{equation}
    \begin{split}
        \mathbf{Q}^T &= \mathbf{X}^T\odot|_{(-1,0)} \mathbf{W}^Q\\
        \mathbf{K}^N &= \mathbf{X}^N\odot|_{(-1,0)} \mathbf{W}^K\\
        \mathbf{V}^N &= \mathbf{X}^N\odot|_{(-1,0)} \mathbf{W}^V
    \end{split}
\label{eq:attention_qkv}
\end{equation}
where $\mathbf{W}^Q$, $\mathbf{W}^K$, and $\mathbf{W}^V$ are learnable parameter kernels (we denote them to belong to {\em query}, {\em key} and {\em value} dense layers, respectively in Figure~\ref{fig:framework}). The operation $(\cdot)\odot|_{(-1,0)}(\cdot)$ signifies {\em Tensor contraction} according to {\em Einstein summation convention} along the specified axis. In Eq.~\ref{eq:attention_qkv}, $(-1,0)$ signifies last and first axis of the first and second tensor, respectively. Therefore, $X\odot|_{(-1,0)}Y = \sum_i X[\dots, i]Y[i, \dots]$. Each of $\mathbf{W}^Q$, $\mathbf{W}^K$, and $\mathbf{W}^V$ is a two-dimensional tensor with $hdim$ columns (last axis).

Next, we compute the attention weight tensor $\mathbf{A}$ between the tweet and news sequence as
\begin{equation}
\label{eq:attention_dot}
    \mathbf{A} = Softmax(\mathbf{Q}^T\odot|_{(-1,-1)}\mathbf{K}^N)
\end{equation}
where $Softmax(X[\dots, i, j]) = \frac{e^{X[\dots, i, j]}}{\sum_j e^{X[\dots, i, j]}}$. Further, to avoid saturation of the softmax activation, we scale each element of $\mathbf{A}$ by $hdim^{-0.5}$ \cite{vaswani2017attention}. 

The attention weight is then used to produce the final encoder feature representation $\mathbf{X}^{T,N}$ by computing the weighted average of $\mathbf{V}^N$ as follows:
\begin{equation}
\label{eq:attention_agg}
    \mathbf{X}^{T,N} = \sum_i \mathbf{V}^N[\dots, i, \dots]\mathbf{A}[\dots, i]
\end{equation}

\framework\ is expected to aggregate the exogenous signal exposed by the sequence of news inputs according to the feature representation of the tweet into $\mathbf{X}^{T,N}$, using the operations mentioned in Eqs. \ref{eq:attention_qkv}-\ref{eq:attention_agg} via tuning the parameter kernels.

{\bf Final prediction.} With $\mathcal{S}^{\text{ex}}$ being represented by the output of the attention framework, we incorporate the features discussed in Section~\ref{subsec:feature_for_retweet} in \framework\ to subsume rest of the signals (see Eq.~\ref{Eq:retweet}). For the two separate modes of retweeter prediction (i.e., static and dynamic), we implement two different variations of \framework.

For the static prediction of retweeters, \framework\ predicts the probability of each of the users $u_1, u_2, \dots, u_n$ to retweet the given tweet with no temporal ordering (see Figure~\ref{fig:framework} (b)). The feature vector $X^{u_i}$ corresponding to user $u_i$ is first normalized and mapped to an intermediate representation using a feed-forward layer. It is then concatenated with the output of the exogenous attention component, $\mathbf{X}^{T,N}$, and finally, another feed-forward layer with sigmoid nonlinearity is applied to compute the probability $P^{u_i}$.

As opposed to the static case, in the dynamic setting \framework\ predicts the probability of every user $u_i$ to retweet within a time interval $t_0+\Delta t_i, t_0+\Delta t_{i+1}$, with $t_0$ being the time of the tweet published and $\Delta t_0=0$. To capture the temporal dependency between predictions in successive intervals, we replace the last feed-forward layer with a Gated Recurrent Unit (GRU), as shown in Figure~\ref{fig:framework} (c). We experimented with other recurrent architectures as well; performance degraded with simple RNN and no gain with LSTM.  

{\bf Cost/loss function.} In both the settings, the task translates to a binary classification problem of deciding whether a given user will retweet or not. Therefore, we use standard {\em binary cross-entropy loss} $L$ to train \framework:
\begin{equation}
\label{eq:loss}
    L = -w\cdot t\log(p) - (1-t)\log(1-p)
\end{equation}
where $t$ is the ground-truth, $p$ is predicted probability ($P^{u_i}$ in static and $P^{u_i}_j$ in dynamic settings), and $w$ is a the weight given to the positive samples to deal with class imbalance.
\section{Experiment Details}
\label{sec:experiments}

\subsection{Dataset}
\label{subsec:data}

% \begin{table}[!t]
% \caption{Performance of different hate speech detection models on our manually annotated test data.}
%     \centering
%     \begin{tabular}{|l|c|c|}
%     \hline
%         {\bf Classifier} & {\bf Macro F1} & {\bf AUC} \\
%     \hline
%         Davidson et al. & 0.59 & 0.847\\
%         Waseem et al. & 0.50 & 0.865\\
%         Pinkesh et al. & 0.49 & 0.5\\
%     \hline
%     \end{tabular}
    
%     \label{tab:hate-detect}
% \end{table}

\begin{SCtable*}
\caption{\small {\bf Statistics of the data crawled from Twitter.} {\em Avg. RT}, {\em Users}, and {\em Users-all} signify average retweets, unique number of users tweeting and the unique number of users engaged in (tweet+retweet) the \#-tag, respectively. 
 JV: {\em jamiaviolence},
 MOTR: {\em MigrantsOnTheRoad},
 TTSV: {\em time\-tosackvadras},
 JUA: {\em jamiaunderattack},
 IBN: {\em IndiaBoycottsNPR},
 ZNBK: {\em ZeeNewsBanKaro},
 SCW: {\em SaluteCoronaWarriors},
 IPIM: {\em IslamoPhobicIndianMedia},
 DR2020: {\em delhiriots2020},
 S4S: {\em Seva4Society},
 PMCF: {\em PMCaresFunds},
 C\_19: {\em COVID\_19},
 HUA: {\em Hindus\_Under\_Attack},
 WP: {\em WarisPathan},
 LE: {\em lockdownextension},
 JCCTV: {\em JamiaCCTV},
 TVI: {\em TrumpVisitIndia},
 PNOP: {\em PutNationOverPublicity},
 DE: {\em DelhiExodus},
 DER: {\em DelhiElectionResults},
 ASMR: {\em amitshahmustresign},
 R4GK: {\em Restore4GinKashmir},
 DV: {\em DelhiViolance},
 SNPR: {\em StopNPR},
 1C4DH: {\em 1Crore4DelhiHindu},
 NV: {\em NirbhayaVerdict},
 NM: {\em NizamuddinMarkaz},
 90DSB: {\em 90daysofshaheenbagh},
 DEM: {\em Demonetisation},
 NHR: {\em NorthDelhiRiots},
 PMP: {\em PMPanuti},
 HLM: {\em HinduLivesMatter},
 CV: {\em ChineseVirus},
 UM: {\em UmarKhalid}.
}
    % \centering
    \scalebox{0.7}{
    \begin{tabular}{|l|c|c|c|c|c|c|c|c|c|c|}
    \hline
    {\bf \#-tags} & JV & MOTR & TTSV & JUA & IBN & ZNBK & SCW & DEM & CV\\
    \hline
    {\bf Tweets} & \(950\) & \(872\) & \(280\) & \(263\) & \(570\) & \(919\) & \(104\) & \(1696\) & \(8\)\\
    {\bf Avg. RT} & \(15.45\) & \(6.69\) & \(8.19\) & \(5.8\) & \(7.87\) & \(9.58\) & \(5.65\) & \(3.46\) & \(0.25\)\\
    % {\bf Duration} & \(16 Feb - 20 Feb\) & \(29Mar-31Mar\) & \(10Mar-10Mar\)  & \(10Feb-11Feb\)  & \(11Mar-13Mar\)  & \(9Feb-11Feb\)  & \(04Apr-09Apr\)  \\
    {\bf Users} & \(743\) & \(641\) & \(138\) & \(215\) & \(333\) & \(751\) & \(53\) &\(607\) & \(7\)\\
    {\bf Users-all} & \(4026\)  & \(2176\) & \(548\) & \(688\) & \(1227\) & \(1940\) & \(225\) &\(4494\) & \(8\)\\
    {\bf \%-Hate} & \(3.78\%\) & \(8.20\%\) & \(1.3\%\) & \(6.06\%\) & \(0.8\%\) & \(7.01\%\) & \(0.0\%\) &\(0.06\%\) & \(0.5\%\)\\
    \hline
    {\bf \#-tags} & IPIM & DR2020 & S4S & PMCF & C\_19 & HUA & WP & NHR & UM\\
    \hline
    {\bf Tweets} & \(4307\) & \(1453\) & \(1087\) & \(1172\) & \(971\) & \(382\) & \(989\) & \(3418\) & \(887\)\\
    {\bf Avg. RT} & \(15.46\) & \(12.23\) & \(13.24\) & \(7.61\) & \(6.38\) & \(7.10\) & \(9.23\) & \(2.89\) & \(3.82\)\\
    % {\bf Duration} & \(03Mar-04Mar\)  & \(26Feb-26Feb\)  & \(31Mar-04Apr\)  & \(30Mar-30Mar\)  & \(13Apr-13Apr\)  & \(02Mar-04Mar\)  & \(20Feb-22Feb\)  \\
    {\bf Users} & \(1181\) & \(1136\) & \(532\) & \(1076\) & \(807\) & \(292\) & \(807\) & \(1316\) & \(439\)\\
    {\bf Users-all} & \(3237\) & \(6051\) & \(4058\) & \(2691\) & \(2593\) & \(1073\) & \(2924\) & \(7251\) & \(2510\)\\
    {\bf \%-Hate} & \(8.42\%\) & \(6.8\%\) & \(1.53\%\) & \(0.8\%\) & \(1.96\%\) & \(10.1\%\) & \(12.07\%\) & \(0.08\%\) & \(0.1\%\)\\
    \hline
    {\bf \#-tags} & LE & JCCTV & TVI & PNOP & DE & DER & ASMR & PMP & \(-\)\\
    \hline
    {\bf Tweets} & \(107\) & \(1045\) & \(339\) & \(555\) & \(542\) & \(843\) & \(959\) & \(1346\) & \(-\)\\
    {\bf Avg. RT} & \(1.85\) & \(12.07\) & \(8.47\) & \(13.24\) & \(9.66\) & \(7.56\) & \(5.01\) & \(4.06\) & \(-\)\\
    % {\bf Duration} & \(09Apr-09Apr\)  & \(17Feb-20Feb\)  & \(13Feb-20Feb\)  & \(03Apr-04Apr\)  & \(28Mar-31Mar\)  & \(11Feb-11Feb\)  & \(03Feb-04Feb\)  \\
    {\bf Users} & \(102\) & \(815\) & \(284\) & \(365\) & \(414\) & \(731\) & \(765\) & \(368\) & \(-\)\\
    {\bf Users-all} & \(138\) & \(4091\) & \(1134\) & \(2146\) & \(1857\) & \(1807\) & \(1807\) & \(2310\) & \(-\)\\
    {\bf \%-Hate} & \(0.0\%\) & \(5.66\%\) & \(2.6\%\) & \(5.71\%\) & \(7.61\%\) & \(3.20\%\) & \(9.94\%\) & \(0.02\%\) & \(-\)\\
    \hline
    {\bf \#-tags} & R4GK & DV & SNPR & 1C4DH & NV & NM & 90DSB & HML & \(-\)\\
    \hline
    {\bf Tweets} & \(949\) & \(1121\) & \(82\) & \(889\) & \(649\) & \(1124\) & \(226\) & \(392\) & \(-\)\\
    {\bf Avg. RT} & \(3.94\) & \(9.004\) & \(10.23\) & \(11.62\) & \(7.61\) & \(8.24\) & \(5.25\) & \(4.82\) & \(-\)\\
    % {\bf Duration} & \(16Mar-18Mar\)  & \(02Mar-04Mar\)  & \(21Feb-22Feb\)  & \(17Mar-18Mar\)  & \(14Mar-18Mar\)  &  \(01Apr-01Apr\)  & \(14Mar-16Mar\) \\
    {\bf Users} & \(492\) & \(948\) & \(64\) & \(770\) & \(546\) & \(843\) & \(188\) & \(145\) & \(-\)\\
    {\bf Users-all} & \(986\) & \(2702\) & \(440\) & \(3045\) & \(1577\) & \(3199\) & \(506\) & \(1396\) & \(-\)\\
    {\bf \%-Hate} & \(2.84\%\) & \(7.37\%\) & \(0.0\%\) & \(0.99\%\) & \(4.67\%\) & \(7.85\%\) & \(12.04\%\) & \(0.12\%\) & \(-\)\\
    \hline
    \end{tabular}}
    \label{tab:tweet_stat}
    \vspace{-5mm}
\end{SCtable*}

We initially started collected data based on topics which led to a tweet corpus spanning across multiple years. To narrow down our time frame and ease the mapping of tweets to news, we restricted our time span from \formatdate{03}{02}{2020} to \formatdate{14}{04}{2020} and made use of trending hashtags. Using Twitter's official API\footnote{\protect\url{https://developer.twitter.com/}}, we tracked and crawled for trending hashtags each day within this duration. Overall, we obtained $31,133$ tweets from $13,965$ users. We also crawled the retweeters for each tweet along with the timestamps. Table \ref{tab:tweet_stat} describes the hashtag-wise detailed statistics of the data. To build the information network, we collected the followers of each user up to a depth of $3$, resulting in a total of $41,151,251$ unique users in our dataset. We also collect the activity history of the users, resulting in a total of $163,042,612$ tweets in our dataset. One should note that the lack of a wholesome dataset (containing textual, temporal, network signals all in one) is the primary reason why we decided to collect our own dataset in the first place.

We also, crawled the online news articles published within this span using the News-please crawler \cite{Hamborg2017}. We managed to collect a total of $683,419$ news articles for this period. After filtering for language, title and date, we were left with $319,179$ processed items. There headlines were used as the source of the exogenous signal.

\subsection{Detecting hateful tweets}
\label{subsec:hate_detect}
We employ three professional annotators who have experience in analyzing online hate speech to annotate the tweets manually. All of these annotators belong to an age group of 22-27 years and are active on Twitter. As the contextual knowledge of real-world events plays a crucial role in identifying hate speech, we ensure that the annotators are well-aware of the events related to the hashtags and topics. Annotators were asked to follow Twitter's policy as guideline for identifying hateful behavior \footnote{\url{https://help.twitter.com/en/rules-and-policies/hateful-conduct-policy}}. We annotated a total of $17,877$ tweets with an inter-annotator agreement of $0.58$ Krippendorf's $\alpha$. The low value of inter-annotator's agreement is at par with most hate speech annotation till date, pointing out the hardness of the task even for human subjects. This further strengthens the need for contextual knowledge as well as exploiting beyond-the-text dynamics. We select the final tags based on majority voting.

Based on this gold-standard annotated data, we train three different hate speech classifiers based on the designs given by Davidson et al. \cite{Davidson2017AutomatedHS} (dubbed as Davidson model), Waseem and Hovy \cite{waseem-hovy-2016-hateful}, and Pinkesh et al. \cite{PinkeshHate}. With an AUC score $0.85$ and macro-F1 $0.59$, the Davidson model emerges as the best performing one. When the existing pre-trained Davidson model was tested on our annotated dataset, it achieved $0.79$ AUC and  $0.48$ macro-F1. This highlights both the limitations of existing hate detection models to capture newer context, as well as the importance of manual annotations and fine-tuning. We use the fine-tuned model to annotate the rest of the tweets in our dataset (\% of hateful tweets for each hashtag is reported in Table~\ref{tab:tweet_stat}). We use the machine-annotated tags for the features and training labels in our proposed models only, while the hate generation models are tested solely on gold-standard data.

Along with the manual annotation and trained hate detection model, we use a dictionary of hate lexicons proposed in \cite{kapoor2018mind}. It contain a total of $209$ words/phrases signaling a possible existence of hatefulness in a tweet. Example of slur terms used in the lexicon include words such as \textit{harami} (bastard), \textit{jhalla} (faggot), \textit{haathi} (elephant/fat). Using the above terms is derogatory and a direct offense. In addition, the lexicon has some colloquial terms such as \textit{mulla} (muslim), \textit{bakar} (gossip), \textit{aktakvadi} (terrorist), \textit{jamai} (son-in-law) which may carry a hateful sentiment depending on the context in which they are used.

\subsection{Hate generation}
\label{subsec:task1_details}

To experiment on our hate generation prediction task, we use a total of $19,032$ tweets(which  have atleast  60 news mapping to it from the time of its posting) coming from $12,492$ users to construct the ground-truth. With an $80:20$ train-test split, there are $611$ hateful tweets among $15,225$ in the training data, whereas $129$ out of $3,807$ in the testing data. To deal with the severe class imbalance of the dataset, we use both upsampling of positive samples and downsampling of negative samples.

With all the features discussed in Section~\ref{sec:hate_generate}, the full size of the feature vector is $3,645$. We experimented with all our proposed models with this full set of features and dimensionality reduction techniques applied to it. We use Principal Component Analysis (PCA) with the number of components
set to \(50\). Also, we conduct experiments selecting $K$-best features ($K=50$) using mutual information.

We implement a total of six different classifiers using Support Vector Machine (with linear and RBF kernel), Logistic Regression, Decision Tree, AdaBoost, and XGBoost \cite{chen2016xgboost}. Parameter settings for each of these are reported in Table~\ref{tab:param-detail}. All of the models, PCA, and feature section are implemented using scikit-learn\footnote{\protect\url{https://scikit-learn.org/stable/}}.

\begin{table}[!t]
\caption{{\bf List of model parameters used for predicting hate generation.} {\em SVM-r} and {\em SVM-l} refer to Support Vector Machine with rbf and linear kernels, respectively. {\em LogReg}: Logistic Regression, {\em Dec-Tree}:  Decision Tree.}
    \centering
    \begin{tabular}{|l|c|}
    \hline
        {\bf Classifier} & {\bf Parameters} \\
    \hline
     
     LogReg & Random state=0\\
     AdaBoost & Random State=1\\
     SVM-r &  Class Weight = `Balanced'\\
     SVM-l & Penalty= l2, Class Weight = `Balanced' \\
     Dec-Tree & Class Weight = `Balanced', Max Depth = 5\\
     XGBoost & eta=0.4, eval metric= `logloss', learning rate=0.0001,\\
    & objective= 'binary:logistic', reg alpha = 0.9\\
    \hline
    \end{tabular}
    \label{tab:param-detail}
    \vspace{-5mm}
\end{table}

\subsection{Retweeter prediction}
\label{subsec:task2_details}

The activity of retweeting, too, shows a skewed pattern similar to hate speech generation. While the maximum number retweets for a single tweet is $196$ in our dataset, the average remains to be $13.10$. We use only those tweets which have more than one retweet and atleast 60 news mapping to it from the time of its posting. With an $80:20$ train-test split, this results in a total of $3,057$ and $765$ samples for training and testing.

For all the Doc2Vec generated feature vectors related to tweets and news headlines, we set the dimensionality to $50$ and $500$, respectively. For \framework, we set the parameter $hdim$ and all the intermediate hidden sizes for the rest of the feed-forward (except the last one generating logits) and recurrent layers to $64$ (see Section \ref{subsec:framework}).

{\bf Hyperparameter tuning of \framework.} For both the settings (i.e, static and dynamic prediction of retweeters), we used mini-batch training of \framework, with both Adam and SGD optimizers. We varied the batch size within $16$, $32$ and $64$, with the best results for a batch size of $16$ for the static mode and $32$ for the dynamic mode. We also varied the learning rates within a range $10^{-4}$ to $10^{-1}$, and chose the best one with learning rate $10^{-2}$ using the SGD optimizer\footnote{\protect{\url{https://www.tensorflow.org/api_docs/python/tf/keras/optimizers/SGD}}} for the dynamic model. The static counterpart produced the best results with Adam optimizer\footnote{\protect{\url{https://www.tensorflow.org/api_docs/python/tf/keras/optimizers/Adam}}} \cite{DBLP:journals/corr/KingmaB14} using default parameters.

To deal with the class imbalance, we set the parameter $w$ in Eq.~\ref{eq:loss} as $w = \lambda(\log C-\log C^{+})$, where $C$ and $C^{+}$ are the counts for total and positive samples, respectively in the training dataset, and $\lambda$ is a balancing constant which we vary from $1$ to $2.5$ with $0.5$ steps. We found the best configurations with
$\lambda=2.0$ and $\lambda=2.5$
for the static and dynamic modes respectively. 

\begin{table*}
\caption{{\bf Evaluation of different classifiers for the prediction of hate generation.} {\em Proc.} signifies different feature selection and label sampling methods, where {\em DS}: downsampling of dominant class, {\em US}: upsampling of dominated class, {\em PCA}: feature dimensionality reduction using PCA, {\em top-K}: selecting top-$K$ features with $K=50$.}
\centering
\scalebox{0.95}{
\begin{tabular}{|l|l|ccc|l|l|ccc|l|l|ccc|}
\hline
Model & Proc. & \multicolumn{1}{c}{Macro-F1} & \multicolumn{1}{c}{ACC} & AUC & Model & Proc. & Macro-F1 & ACC & AUC & Model & Proc & \multicolumn{1}{c}{Macro-F1} & \multicolumn{1}{c}{ACC} & AUC \\ \hline
\multirow{6}{*}{\begin{tabular}[c]{@{}l@{}}SVM\\ linear\end{tabular}} & None & \(0.52\)  &\(0.94\)  & \(0.52\)  & \multirow{6}{*}{\begin{tabular}[c]{@{}l@{}}SVM\\ rbf\end{tabular}} & None & \(0.55\) & \(0.88\) & \(0.61\)  & \multirow{6}{*}{LogReg} & None & \(0.50\)  & \(0.96\)  &  \(0.503\) \\
 & DS & \(0.63\) & \(0.73\) & \(0.63\) &  & DS & \(0.62\) & \(0.70\) & \(0.64\) &  & DS & \(0.64\) &  \(0.79\)&\(0.629\)  \\
 & US+DS &\(0.44\)  & \(0.64\) &  \(0.63\)&  & US+DS &\(0.46\)  & \(0.69\) &\(0.66\)  &  & US+DS &\(0.47\)  &\(0.72\)  &  \(0.63\)\\
 & PCA & \(0.55\) &\(0.90\)  &  \(0.59\)&  & PCA & \(0.48\) & \(0.71\) & \(\bf 0.68\) &  & PCA &\(0.49\)  & \(0.97\) &  \(0.50\)\\
 & top-K & \(0.53\) & \(0.84\) &\(0.63\)  &  & top-K &\(0.50\)  & \(0.79\) & \(0.62\) &  & top-K & \(0.49\) &\(0.97\)  & \(0.50\) \\ 
 %\cline{2-5} \cline{7-10} \cline{12-15} 
%  & PCA+DS &\(0.57\)  & \(0.71\) & \(0.57\) &  & PCA+DS &\(0.63\)   &  \(0.68\)&\(0.66\)  &  &PCA+DS &\(0.60\)  & \(0.78\) & \(0.59\) \\ 
\hline
\multirow{6}{*}{\begin{tabular}[c]{@{}l@{}}Dec-\\ Tree\end{tabular}} & None &\(0.51\)  &\(0.79\)  & \(0.64\) & \multirow{6}{*}{\begin{tabular}[c]{@{}l@{}}Ada-\\ Boost\end{tabular}} & None &\(0.49\)  &\(0.97\)  &  \(0.49\)& \multirow{6}{*}{XGB} & None &  \(0.53\)& \(0.97\) & \(0.52\) \\
 & DS &\(\bf 0.65\)  & \(0.74\) & \(0.66\) &  & DS &\(0.62\)  &\(0.77\)  & \(0.61\) &  & DS & \(0.57\) & \(0.76\) &\(0.566\)  \\
 & US+DS &\(0.45\)  &\(0.67\)  &\(0.61\)  &  & US+DS &\(0.44\)  &\(0.63\)  &\(\bf 0.68\)  &  & US+DS &\(0.44\)  &\(0.66\)  &\(0.62\)  \\
 & PCA & \(0.46\) &  \(0.68\)&\(0.65\)  &  & PCA & \(0.50\) &\(0.97\)  &\(0.50\)  &  & PCA &\(0.51\)  &\(0.96\)  &  \(0.51\)\\
 & top-K &\(0.53\)  &  \(0.84\)& \(0.63\) &   & top-K &\(0.49\)  &  \(0.97\)& \(0.50\) &  & top-K &\(0.49\)  & \(0.97\) & \(0.50\) \\ 
 %\cline{2-5} \cline{7-10} \cline{12-15} 
%  & PCA+DS & \(0.60\) &\(0.66\)  &  \(0.63\)&  & PCA+DS &\(0.61\)  & \(0.78\) &\(0.59\)  &  & PCA+DS &  \(0.56\)&\(0.73\)  &\(0.55\)  \\ 
 \hline
\end{tabular}}
\label{tab:task1-full}
\vspace{-5mm}
\end{table*}
\section{Baselines and ablation variants}
\label{sec:baselines}
In the absence of external baselines for predicting hate generation probability due to the problem's novelty, we explicitly rely on ablation analyses of the models proposed for this task. For retweet dynamics prediction, we implement $5$ external baselines and two ablation variants of \framework. Since information diffusion is a vast subject, we approach it from two perspectives -- one is the set of rudimentary baselines (SIR, General Threshold), and the other is the set of recently proposed neural models.
\subsection{Rudimentary baselines}
% \begin{enumerate}
\textbf{SIR} \cite{SIR_Original}: The Susceptible-Infectious-Recovered (Removed)   is one of the earliest predictive models for contagion spread. Two parameters govern the model -- transmission rate and recovery rate, which dictate the spread of contagion (retweeting in our case) along with a social/information network.

\textbf{Threshold Model} \cite{kempe2003maximizing}: This model assumes that each node has threshold inertia chosen uniformly at random from the interval \([0,1]\). A node becomes active if the weighted sum of its active neighbors exceeds this threshold.
% \end{enumerate}
\subsection{Feature engineering based baselines}
Using the same feature set as described in Section \ref{subsec:feature_for_retweet}, we employ four classifiers -- Logistic Regression, Decision Tree, Linear SVC, and Random Forest (with 50 estimators). All of these models are used for the static mode of retweet prediction only. Features representing exogenous signals are engineered in the same way as described in Section \ref{subsec:exogen_feature}.

\subsection{Neural network based  baselines}
To overcome the feature engineering step involving combinations of topical, contextual, network, and user-level features, neural methods for information diffusion have gained popularity. While these methods are all focused on determining only the next set of users, they are still important to measure the diffusion performance of \framework.

\textbf{TopoLSTM} \cite{TopoLSTM}:   It is one of the initial works to consider recurrent models in generating the next user prediction probabilities. The model converts the cascades into dynamic DAGs (capturing the temporal signals via node ordering). The sender-receiver based RNN model captures a combination of active node's static score (based on the history of the cascade), and a dynamic score (capturing future propagation tendencies). 
    
\textbf{FOREST} \cite{Yang2019MultiscaleID}: It aims to be a unified model, performing the microscopic and the macroscopic cascade predictions combining reinforcement learning (for macroscopic) with the recurrent model (for microscopic). By considering the complete global graph, it performs graph sampling to obtain the structural context of a node as an aggregate of the structural context of its one or two hops neighbors. In addition, it factors the temporal information via the last $m$ seen nodes in the cascade. 
% The sender-receiver based RNN model is enhanced by the action-state model of RL, to give superior performance for both the tasks. It is worth noticing that largest graph the model was tested on consisted of $\approx 23k$ nodes, while even after sampling our smallest graph has $\approx 150k$ nodes. This six-fold increase in data, coupled with limited computing power, drastically reduced the performance of FOREST, as it tries to predict next user probabilities against all $150k$ nodes.
    
\textbf{HIDAN} \cite{Wang2019HierarchicalDA}: It does not explicitly consider a global graph as input. Any information loss due to the absence of a global graph is substituted by temporal information utilized in the form of ordered time difference of node infection. 
% The model uniquely captures the non-sequential nature of information diffusion, by employing a two-tier attention mechanism. The first attention system, builds the structural context of a user from the historical inter-user interactions. This context is aggregated into user's own context. Meanwhile, at cascade-level the temporal information is used, along with the updated user context to predict the next node. 
Since HIDAN does not employ a global graph, like TopoLSTM, it too uses the set of all seen nodes in the cascade as candidate nodes for prediction.
% \end{enumerate}

\subsection{Ablation models}
\label{subsec:ablation}
% In the absence of external baselines for predicting hate generation probability due to the problem's novelty, 
We exercise extensive feature ablation to examine the relative importance of different feature sets. Among the six different algorithms we implement for this task, along with different sampling and feature reduction methods, we choose the best performing model for this ablation study. Following Eq.~\ref{Eq:tweet}, we remove the feature sets representing $\mathcal{H}_{i,t}$, $\mathcal{S}^\text{ex}$, $\mathcal{S}^\text{en}$, and $\mathcal{T}$ (see Section~\ref{sec:hate_generate} for corresponding features) in each trial and evaluate the performance. 

To investigate the effectiveness of the exogenous attention mechanism for predicting potential retweeters, we remove this component and experiment on static as well as the dynamic setting of \framework.

\section{Evaluation}
\label{sec:evamjoation}
Evaluation of classification models on highly imbalanced data needs careful precautions to avoid classification bias. We use multiple evaluation metrics for both the tasks: macro averaged F1 score (macro-F1), area under the receiver operating characteristics (AUC), and binary accuracy (ACC). As the neural baselines tackle the problem of retweet prediction as a ranking task, we improvise the evaluation of \framework\ to make it comparable with these baselines. We rank the predicted probability scores ($P^{u_i}$ and $P^{u_i}_j$ in static and dynamic settings, respectively) and compute mean average precision at top-$k$ positions (MAP@$k$) and binary hits at top-$k$ positions (HITS@$k$).

\begin{table}[!t]
\caption{{\bf Feature ablation for Decision Tree with downsampling for predicting hate generation.} At each trial, we remove features representing signals -- $\mathcal{H}_{i,t}$ ({\em All $\setminus$ History}), $\mathcal{S}^\text{ex}$ ({\em All $\setminus$ Endogen}), $\mathcal{S}^\text{en}$ ({\em All $\setminus$ Exogen}), and $\mathcal{T}$ ({\em All $\setminus$ Topic}). See Eq.~\ref{Eq:tweet} and Section~\ref{sec:hate_generate} for details of the signals and features, respectively.}
    \centering
    \begin{tabular}{|l|c|c|c|}
    \hline
        {\bf Features} & {\bf Macro-F1} & {\bf ACC} & {\bf AUC} \\
    \hline
        All & \(0.65\)&\(0.74\) &\(0.66\) \\
        All $\setminus$	 History &\(0.56\) &\(0.59\) &\(0.64\) \\
        All $\setminus$ Endogen &\(0.61\) &\(0.68\) &\(0.64\) \\
        All $\setminus$ Exogen &\(0.56\) &\(0.58\) &\(0.66\) \\
        All $\setminus$ Topic &\(0.65\) &\(0.74\) & \(0.66\)\\
    \hline
    \end{tabular}
\label{tab:task1-ablation}
\vspace{-5mm}
\end{table}

\hspace{3cm}
\begin{figure}[!t]
    \centering
    \includegraphics[width=0.7\columnwidth]{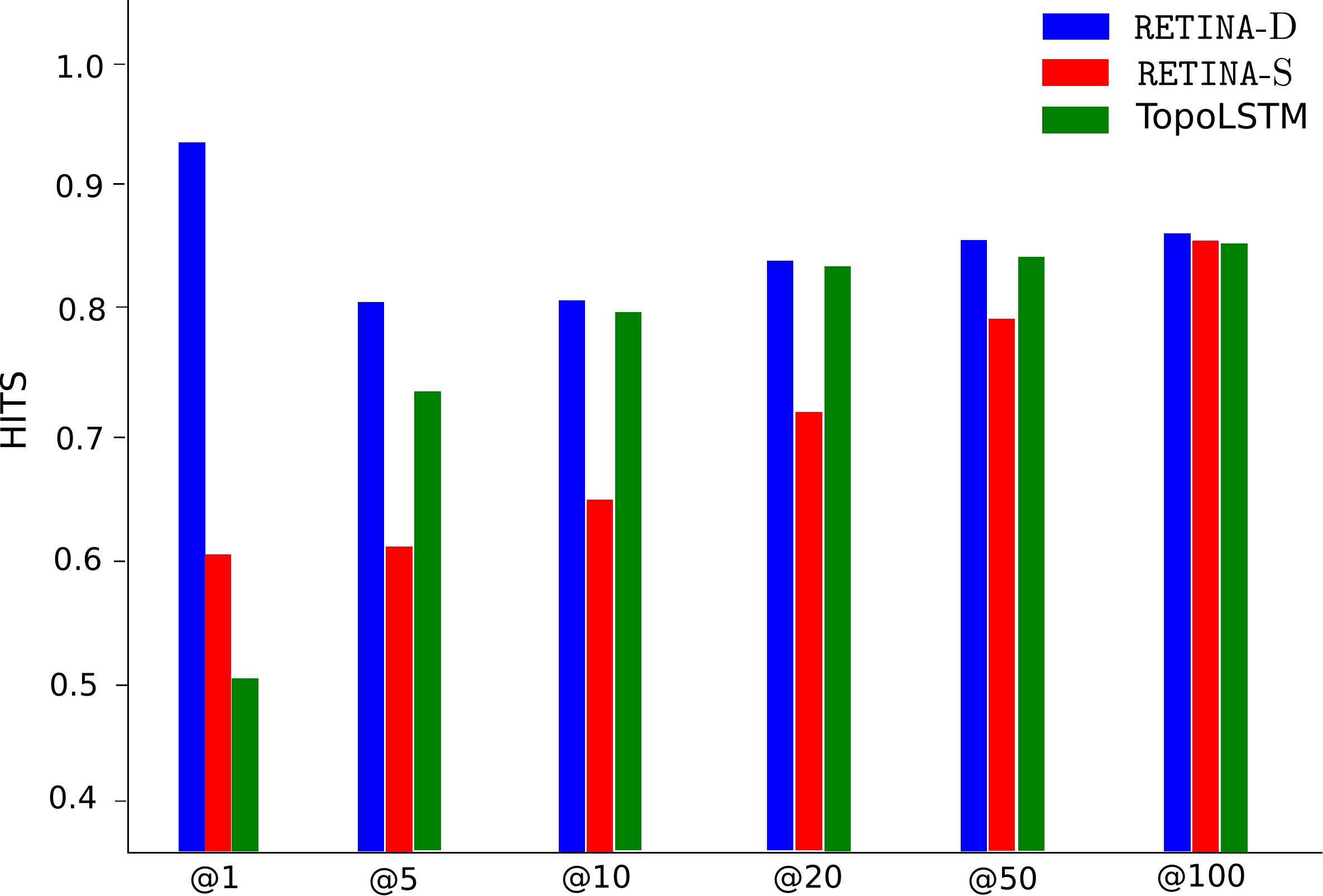}
    \caption{HITS@$k$ of \framework-D, \framework-S, and TopoLSTM for retweeter prediction with $k=1,5,10,20,50,$ and $100$.}
    \label{fig:hits_plot}
    \vspace{-5mm}
\end{figure}
\vspace{-0.5cm}
\subsection{Performance in predicting hate generation}
\label{subsec:task1-evaluation}

Table~\ref{tab:task1-full} presents the performances of all the models  to predict the probability of a given user posting a hateful tweet using a given hashtag. It is evident from the results that, all six models suffer from the sharp bias in data; without any class-specific sampling, they tend to lean towards the dominant class (non-hate in this case) and result in a low macro-F1 and AUC compared to very high binary accuracy. SVM  with rbf-kernel outperforms the rest when no upsampling or downsampling is done, with a macro-F1 of $0.55$ (AUC $0.61$).

{\bf Effects of sampling.} Downsampling the dominant classes result in a substantial leap in the performance of all the models. The effect is almost uniform over all the classifiers except XGBoost. In terms of macro-F1, Decision Tree sets the best performance altogether for this task as $\mathbf{0.65}$. However, the rest of the models lie in a very close range of $0.62$-$0.64$ macro-F1.

While the downsampling performance gains are explicitly evident, the effects of upsampling the dominated class are less intuitive. For all the models, upsampling deteriorates macro-F1 by a large extent, with values in the range $0.44$-$0.47$. However, the AUC scores improve by a significant margin for all the models with upsampling except  Decision Tree. AdaBoost achieves the highest AUC  of $\mathbf{0.68}$ with upsampling.

{\bf Dimensionality reduction of feature space.} Our experiments with PCA and $K$-best feature selection by mutual information show a heterogeneous effect on different models. While the only SVM with linear kernel shows some improvement with PCA over the original feature set, the rest of the models observe considerable degradation of macro-F1. However, SVM with rbf kernel achieves the best AUC  of $0.68$ with PCA. With top-$K$ best features, the overall gain in performance is not much significant except Decision Tree.

We also experiment with combinations of different sampling and feature reduction methods, but none of them achieve a significant gain in performance. 

{\bf Ablation analysis.} We choose Decision Tree with down-sampling of dominant class as our best performing model (in terms of macro-F1 score) and perform ablation analysis. Table~\ref{tab:task1-ablation} presents the performance of the model with each feature group removed in isolation, along with the full model. Evidently, for predicting hate generation, {\bf features representing exogenous signals and user activity history are most important}. Removal of the feature vector signifying trending hashtags, which represent the endogenous signal in our case, also worsens the performance to a significant degree.

\begin{table}[!t]
\caption{{\bf Performance of  \framework\ and other baselines  for retweeter prediction.} \framework-S and \framework-D correspond to static and dynamic prediction settings, respectively. \textsuperscript{$\dagger$} symbolizes models without exogenous signal. 
% {\em Gen.Thresh.} corresponds to the General Threshold model for diffusion prediction.
}
    \centering
    \scalebox{0.8}{
    \begin{tabular}{|l|ccccc|}
    \hline
        {\bf Model} & {\bf Macro-F1} & {\bf ACC} & {\bf AUC} & {\bf MAP@20} & {\bf HITS@20} \\
    \hline
        Logistic Regression &\(0.70\) &\(0.96\) & \(0.79\)& - & - \\
        Logistic Regression\textsuperscript{$\dagger$} &\(0.49\) &\(0.93\) & \(0.50\)& - & - \\
        Decision Tree &\(0.68\) &\(0.95\) & \(0.78\)& - & - \\ 
        Decision Tree\textsuperscript{$\dagger$} &\(0.54\) &\(0.92\) & \(0.54\)& - & - \\ 
        Random Forest &\(0.66\) &\(0.97\) & \(0.67\)& - & - \\
        Random Forest\textsuperscript{$\dagger$} &\(0.52\) &\(0.93\) & \(0.52\)& - & - \\
        Linear SVC\textsuperscript{$\dagger$} &\(0.49\) &\(0.91\) & \(0.50\)& - & - \\
        \framework-S &\(0.70\) &\(0.97\) & \(0.73\)& \(0.57\)&\(0.74\) \\
        \framework-S\textsuperscript{$\dagger$} &\(0.65\) & \(0.93\)& \(0.74\)& \(0.56\)&\(0.76\) \\
        %LReg-D &\(0.94\) & \(1.0\)&\(98.3\) & & \\
        \framework-D &\(\bf 0.89\) & \(0.99\)&\(\bf 0.86\) &\(\bf 0.78\) &\(\bf 0.88\) \\
        \framework-D\textsuperscript{$\dagger$} &\(0.87\) &\(0.99\) &\(0.798\) &\(0.69\) &\(0.80\) \\
        FOREST & - & - & - &\(0.51\) & \(0.64\)\\
        HIDAN & - & - & - &\(0.05\) & \(0.05\)\\
        TopoLSTM & - & - & - & \(0.60\)&\(0.83\) \\
        SIR &\(0.04\) & - & - & - & - \\
        Gen.Thresh. & \(0.04\)& - & - & - & - \\
    \hline
    \end{tabular}}
    \label{tab:task2-all}
    \vspace{-3mm}
\end{table}

\subsection{Performance in retweeter prediction}
\label{subsec:task2-eval}

Table~\ref{tab:task2-all} summarizes the performances of the competing models for the retweet prediction task. Here again, binary accuracy presents a very skewed picture of the performance due to class imbalance. While \framework\ in dynamic setting outperforms the rest of the models by a significant margin for all the evaluation metrics, TopoLSTM emerges as the best baseline in terms of both MAP@20 and HITS@20. 

In Figure~\ref{fig:hits_plot}, we compare \framework\ in static and dynamic setting with TopoLSTM in terms of HITS@$k$ for different values of $k$. For smaller values of $k$, \framework\ largely outperforms TopoLSTM, in both dynamic and static setting. However, with increasing $k$-values, the three models converge to very similar performances.

\begin{figure}
    \centering
    \includegraphics[width=0.7\columnwidth]{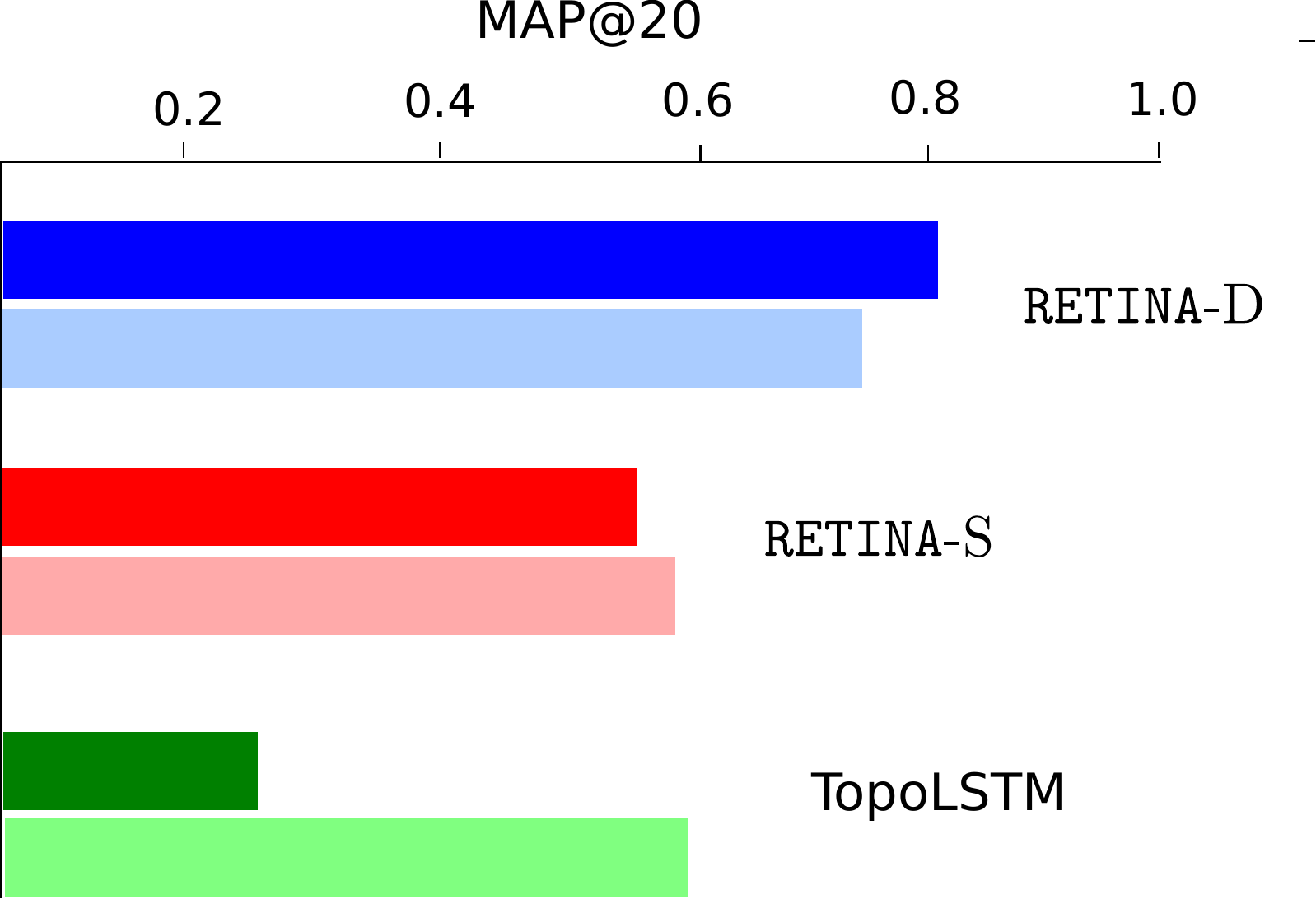}
    \caption{Comparison of \framework\ in static (red; \framework-S) and dynamic (blue; \framework-D) setting  with TopoLSTM (green) to predict potential retweeters when the root tweet is -- hateful (dark shade) vs  non-hate (lighter shade).  }
    \label{fig:h-vs-nh-retweet}
    \vspace{-3mm}
\end{figure}

Figure~\ref{fig:h-vs-nh-retweet} provides an important insight regarding the retweet diffusion modeling power of our proposed framework \framework. Our best performing baseline, TopoLSTM largely fails to capture the different diffusion dynamics of hate speech in contrast to non-hate (MAP@20 $0.59$ for non-hate vs. $0.43$ for hate). On the other hand, \framework\ achieves MAP@20 scores $0.80$ and $0.74$  in dynamic ($0.54$ and $0.56$ in static) settings to predict the retweet dynamics for hate and non-hate contents, respectively. One can readily infer that our well-curated feature design by incorporating hate signals along with the endogenous, exogenous, and topic-oriented influences empowers \framework\ with this superior expressive power.

Among the traditional  baselines, Logistic Regression gives comparable Macro F1-score to the best static model; however, owing to memory limitations it could not be trained on news set larger than 15 per tweet. Similarly, SVM based models could not incorporate even 15 news items per tweet (memory limitation). Meanwhile, an ablation on news size gave best results at 60 for both static and dynamic models. 

\begin{figure}
    \centering
    \includegraphics[width=\columnwidth]{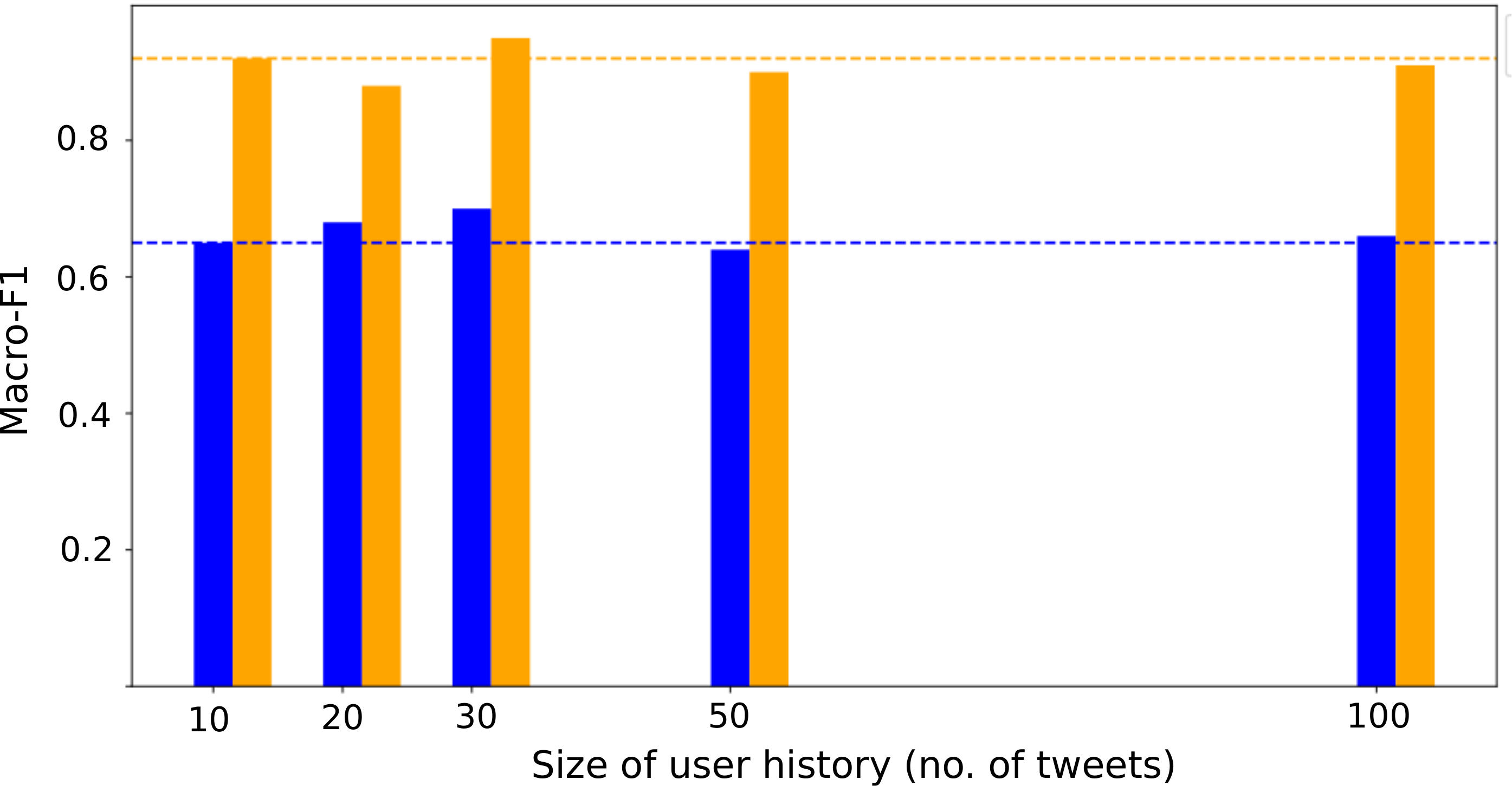}
    \caption{Variation in performance of \framework\ when we consider different number of tweets posted by a user as its tweeting history. Blue and yellow bars represent the static and dynamic modes of retweet prediction, respectively.\\}
    \vspace{-5mm}
    \label{fig:user-hist-vary}
\end{figure}

We find that the contribution of the exogenous signal(i.e the news items) plays a vital role in retweet prediction, much similar to our findings in Table~\ref{tab:task1-ablation} for predicting hate generation. With the exogenous attention component removed in static as well as dynamic settings (\framework-S\textsuperscript{$\dagger$} and \framework-D\textsuperscript{$\dagger$}, respectively, in Table~\ref{tab:task2-all}), performance drops by a significant margin. However, the performance drop is more significant in \framework-D\textsuperscript{$\dagger$} for ranking users according to retweet probability (MAP@$k$ and HITS@$k$). The impact of exogenous signals on Macro-F1 is more visible in the traditional models.

\begin{figure}[!t]
    \centering
    \includegraphics[width=\columnwidth]{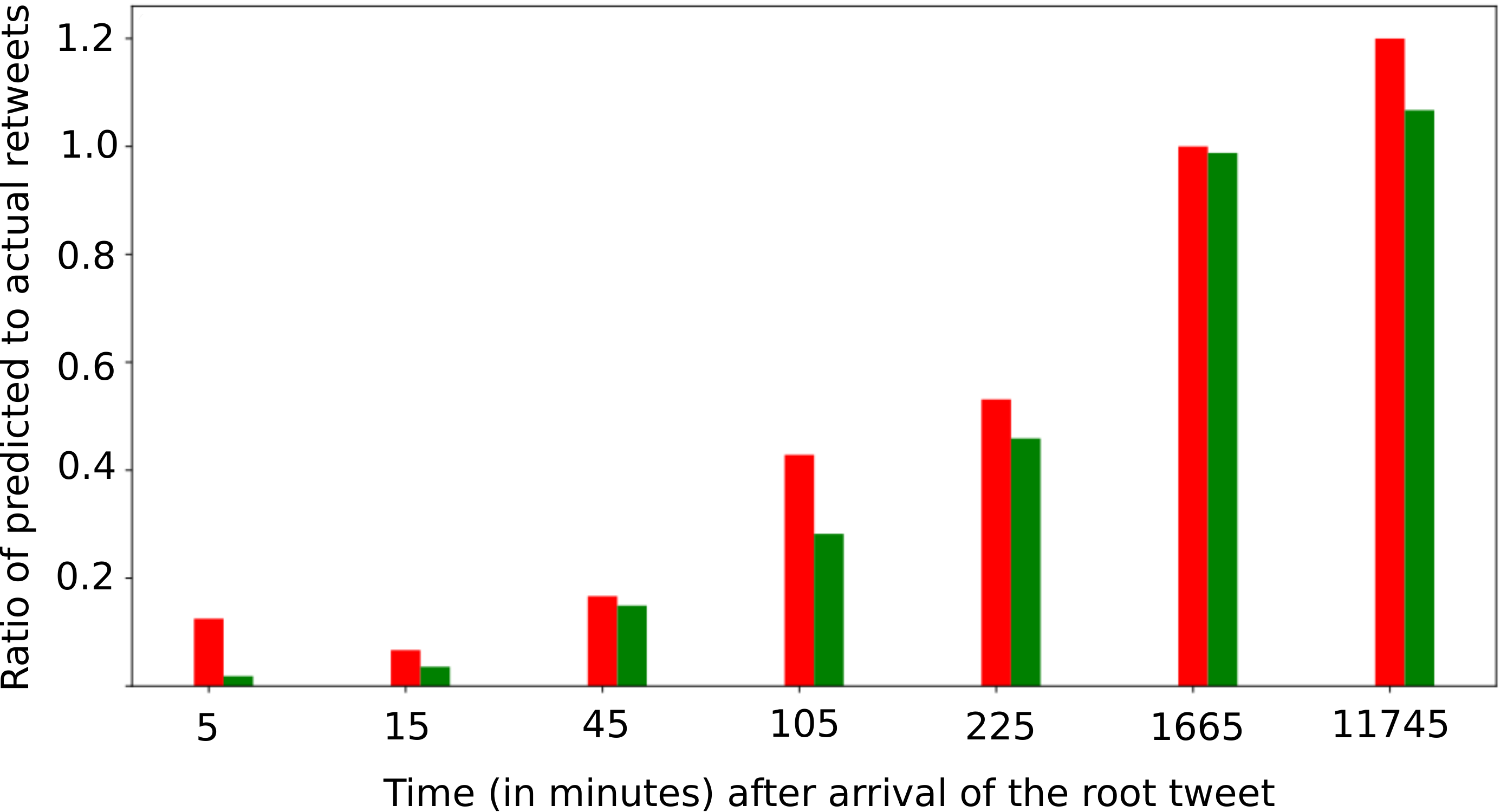}
    \caption{Ratio of the number of predicted to actual retweets arrived within different time-windows after the arrival of the root tweet. Red and green bars correspond to hateful vs non-hate root tweets. Counts of actual and predicted retweets are taken between each successive time instances along the horizontal axis.}
    \vspace{-5mm}
    \label{fig:pred_vs_time}
\end{figure}

\begin{figure}
    \centering
    \includegraphics[width=\columnwidth]{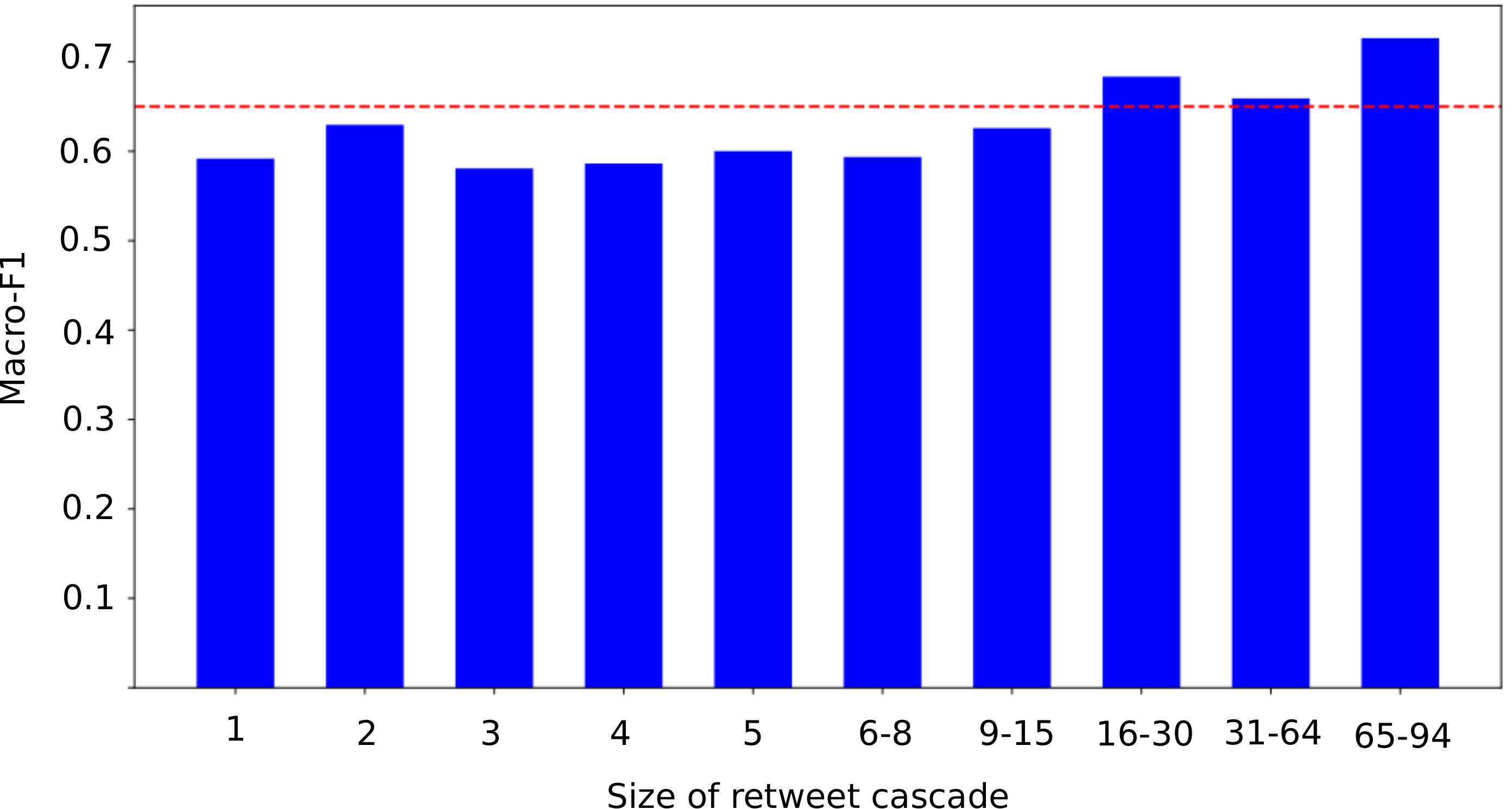}
    \caption{Variation in performance of \framework-S with cascade size. The red dashed line signifies the overall macro-F1.}
    \label{fig:static_cas_size}
    \vspace{-5mm}
\end{figure}

To observe the performance of \framework\ more closely in the dynamic setting, we analyse its performance over successive prediction intervals. Figure~\ref{fig:pred_vs_time} shows the ratio between the predicted and the actual number of retweets arrived at different intervals. As clearly evident, the model tends to be nearly perfect in predicting new growth with increasing time. High error rate at the initial stage is possibly due to the fact that the retweet dynamics remains uncertain at first and becomes more predictable as increasing number of people participate over time. A similar trend is observed when we compare the performance of \framework\ in static setting with varying size of actual retweet cascades.   Figure~\ref{fig:static_cas_size} shows that \framework-S performs better with increasing size of the cascade. 

In addition, we also vary the number of tweets posted by a user.  Figure~\ref{fig:user-hist-vary} shows that the performance of \framework\ in both static and dynamic settings increases by varying history size from $10$ to $30$ tweets. Afterward, it either drops or remains the same.

\section{Limitations of the Current Study}
\label{sec:limitations}
Our attempt to model the genesis and propagation of hate on Twitter brings forth various limitations posed by the problem itself as well as our modeling approaches. We explicitly cover such areas to facilitate the grounds of future developments.

\subsection{Partial observation of hate diffusion}
We have considered the propagation of hateful behavior via retweet cascades only. In practice, there are multiple other forms of diffusion present, and retweet only constitutes a subset of the full spectrum. Users susceptible to hateful information often propagate those via new tweets. Hateful tweets are often counteracted with hate speech via reply cascades. Even if not retweeted, replied, or immediately influencing the generation of newer tweets, a specific hateful tweet can readily set the audience into a hateful state, which may later develop repercussions. Identification of such influences would need intricate methods of Natural Language Processing techniques, adaptable to the noisy nature of Twitter data.

\subsection{Dynamic and subjective nature of hate speech}
As already discussed, online hate speech is vastly dynamic in nature, making it difficult to identify. Depending on the topic, time, cultural demography, target group, etc., the signals of hate speech change. Thus models like \framework\, which explicitly uses hate-based features to predict the popularity, need updated signaling strategy. However, this drawback is only evident if one intends to perceive such endeavors as a simple task of retweet prediction only. We, on the other hand, focus on the retweet dynamics of hateful vs. non-hateful contents which presumes the signals of hateful behavior to be well-defined.

\section{Conclusion}
\label{sec:conclusion}
The majority of the existing studies on online hate speech focused on hate speech detection, with a very few seeking to analyze the diffusion dynamics of hate on large-scale information networks. We bring forth the very first attempt to predict the initiation and spread of hate speech on Twitter. Analyzing a large Twitter dataset that we crawled and manually annotated for hate speech, we identified multiple key factors (exogenous information, topic-affinity of the user, etc.) that govern the dissemination of hate. 

Based on the empirical observations, we developed multiple supervised models powered by rich feature representation to predict the probability of any given user tweeting something hateful. We proposed \framework, a neural framework exploiting extra-Twitter information  (in terms of news) with attention mechanism for predicting potential retweeters for any given tweet. Comparison with multiple state-of-the-art models for retweeter prediction revealed the superiority of \framework\ in general as well as for predicting the spread of hateful content in particular. 

With specific focus of our work being the generation and diffusion of hateful content, our proposed models rely on some general textual/network-based features as well as features signaling hate speech. A possible future work can be to replace hate speech with any other targeted phenomenon like fraudulent, abusive behavior, or specific categories of hate speech. However, these hate signals require a manual intervention when updating the lexicons or adding tropical hate tweets to retrain the hate detection model. While the features of the end-to-end model appear to be highly engineered, individual modules take care of respective preprocessing.

In this study, the mode of hate speech spread we primarily focused on is via retweeting, and therefore we restrict ourselves within textual hate. However, spreading hateful contents packaged by an image, a meme, or some invented slang are some new normal of this age and leave the space for future studies.

\bibliography{ref}
\bibliographystyle{IEEEtran}
\end{document}